\documentclass[usegraphicx,usenatbib,useAMS]{mn2e}

\newcommand{\beq}{\begin{equation}}
\newcommand{\eeq}{\end{equation}}
\newcommand{\be}{\begin{equation}}
\newcommand{\ee}{\end{equation}}
\newcommand{\bbtheta}{\mbox{\boldmath $\theta$}}
\newcommand{\lnL}{\ln{\cal L}}
\def\gap{\;\rlap{\lower 2.5pt
 \hbox{$\sim$}}\raise 1.5pt\hbox{$>$}\;}
\def\lap{\;\rlap{\lower 2.5pt
   \hbox{$\sim$}}\raise 1.5pt\hbox{$<$}\;}

\begin{document}

\title[Dark halo properties]{Dark halo properties from
rotation curves} 
\author[Jimenez, Verde \& Oh] 
{Raul Jimenez$^1$, Licia Verde$^{1,2}$ and S. Peng Oh$^3$ \\ 
$^1$Department of Physics and Astronomy, Rutgers University, 
136 Frelinghuysen Road, Piscataway, NJ 08854--8019, USA.\\
$^2$Princeton University Observatory, Princeton, NJ 08544, USA.\\ 
$^3$Theoretical Astrophysics, Mail Code 130-33, Caltech, Pasadena, 
CA 91125, USA.}

\maketitle

\begin{abstract}
  We study a large set of high spatial resolution optical rotation curves of
  galaxies with the goal of determining the model parameters for a disk
  embedded within a cold dark matter (CDM) halo that we model either with a
  Navarro, Frenk \& White (NFW) profile or pseudo-isothermal profile. We show
  that parameter degeneracies present in lower resolution data are lifted at
  these higher resolutions. 34\% of the galaxies do not have a meaningful fit
  when using the NFW profile and 32\% when using the pseudoisothermal profile,
  however only 14\% do not have a meaningful fit in either model.  In both
  models we find correlations between the disk baryon fraction $f_d$ and the
  spin parameter of the halo $\lambda'$, between $f_d$ and the dark halo mass
  $M_{200}$, and between $M_{200}$ and the concentration parameter $c$. We
  show that the distribution of the concentration parameters $c$, for a NFW
  halo, is in good agreement with CDM predictions; no significant galaxy
  population is found with very low values of $c$. The overall distribution of
  $\lambda'$ is in good agreement with theoretical predictions from
  hierarchical tidal torque theory. The whole sample is also well fitted by a
  pseudo-isothermal dark halo with a core, but the size of the core is rather
  small (6\% of the virial radius or smaller; for 70\% of the sample the core
  size is less than 2 kpc). Thus we conclude that the profile of dark matter
  is steep ($r^{-1}$ or steeper) down to this radius; large dark matter cores
  (and therefore very low dark matter central densities) seem to be excluded.
  LSBs tend to have higher values of $\lambda'$ for a given $f_d$ and lower
  values of $c$ for a given mass than HSBs.  In an appendix we give some
  useful formula for pseudo-isothermal profile halos and discuss in detail the
  issue of parameter degeneracies.
\end{abstract}

\begin{keywords}
cosmology: theory --- galaxies: formation --- galaxies: spiral ---
galaxies: kinematics and dynamics
\end{keywords}

\section{Introduction}

The Cold Dark Matter (CDM) paradigm for structure formation has proved
remarkably successful in explaining the observed large-scale properties of the
universe, such as the abundance and clustering of galaxies and clusters
\citep{Peacock+01,Verde+2dF01,Lahav+2dF01}, the statistical properties of the
Ly$\alpha$ forest (e.g., \citet{Croft+01}), and the power spectrum of the
cosmic microwave background anisotropies (e.g., \citet{Jaffe+01}). However, a
number of puzzling discrepancies remain when CDM predictions are extrapolated
to small scales. Chief among them are: (i) The substructure problem. CDM
over-predicts the number of satellites around a Milky-Way sized galaxy by an
order of magnitude \citep{KKVP99,MGGLQST99}. (ii) The density profile problem.
Observed rotation-curves of dwarf and lower surface brightness galaxies
suggest that the inner regions have a constant density core rather than the
density cusp predicted by CDM (e.g., \citet{MB98,MGGLQST99,DB00}).

Furthermore, numerical SPH simulations of disk formation within dark 
halos fail to  match the zero-point of the observed Tully-Fisher. One possible
solution is to lower the central concentration of the dark matter 
halos \citep{NS00}.
 
These problems have stimulated numerous proposed astrophysical solutions, as
well as modifications of the fundamental CDM paradigm itself (e.g.,
\citet*{SS00,KL00,CDW96,BOT01,Goodman00,HBG00,Cen01}).

In this paper we once again examine the density profile problem by fitting CDM
models to observed rotation curves of spiral galaxies.  Our study differs from
previous work in two respects. Firstly, we use optical rotation curves rather
than the HI rotation curves used in previous studies which are plagued by beam
smearing \citep{SMT00,vanBRDB00}. Even if the effects of beam smearing are
neglected, the relatively large errors and limited spatial sampling of HI
rotation curves imply that they cannot be used to discriminate between
constant density cores and $r^{-1}$ cusps \citep*{vanBS01}. By contrast, the
optical rotation curves we use are free from beam smearing, have smaller
errors and higher spatial resolution. We show that, with this superior data,
various parameter degeneracies present when fitting HI rotation curves can be
lifted.  This allows us to distinguish between core and cusp-like inner
profiles.  Secondly, while many studies have focused on relatively small
samples of dwarfs (but see \citet{Navarro98}), we use a large sample (400
galaxies) spanning a wide range in luminosity and surface brightness.

Our goal is to determine the best fitting model parameters for a disk within
the CDM profile proposed by Navarro Frenk and White (1997; NFW) and within a
pseudo-isothermal profile halo and study the distributions of the recovered
disk parameters.

We find that the NFW profile provides a good fit to 66\% of the galaxies in
the sample, with a distribution of recovered concentration parameters and spin
parameters broadly consistent with that predicted by CDM numerical
simulations. When the sample is fitted with an isothermal profile with a core,
68\% of the galaxies are well fitted whithin this model and the best fit
parameters favor cores with small sizes (below 6\% of the dark halo virial
radius).  Many rotation curves that have no meaningful fit in one dark matter
profile are well fitted by the other. There is no meaningful fit in either
model for only for 14\% of the sample. This is consistent with the inner dark
matter profiles being steep (slope of -1 or steeper) down to radii that are
few percent of the virial radius.

In both models the recovered baryonic mass to light ratios are broadly in
agreement with predictions from synthetic stellar populations.

This paper is organized as follows.  In section 2 we present the set of
rotation curves we analyze and the two models we use to fit the data
(exponential disk embedded in a dark matter halo with a NFW profile or a
pseudo-isothermal profile). These two models have four free parameters, but
since for most of the galaxies the scale length of the disk is given, we
report the analysis for four free parameters in appendix II and present in the
main text the results for the fit with three free parameters.  In section 3 we
illustrate and discuss the correlations we find between the best fit
parameters for the two models. We discuss the general implications of these
findings in section 4 where we also compare the derived baryonic mass-to-light
ratios with the range allowed by stellar populations.  Finally, we conclude
and summarize our results in section 5.  In appendix I we develop the
expression for the rotation curve of an exponential disk embedded in a
pseudoisothermal halo. In appendix III we study the degeneracies among the
parameters of the models.
 
\section{Rotation Curves}
We study a large set of observed rotation curves of galaxies with the aim of
determining the model parameters for a disk within a NFW \citep*{NFW97} profile
halo and within a pseudo-isothermal profile halo (i.e., an isothermal sphere
with a constant density core).  The choice of a pseudo-isothermal halo is
motivated by recent claims in the literature that the dark matter profile of
low--surface brightness galaxies (LSB, e.g., \citet{BMBR01}) and some
high--surface brightness galaxies (HSB, e.g., \citet*{Salucci01,BS01,SB00}) is
better represented by a flat core rather than the steeper profile found in CDM
N-body simulations.

\begin{figure*}
\includegraphics[width=16cm,height=14cm]{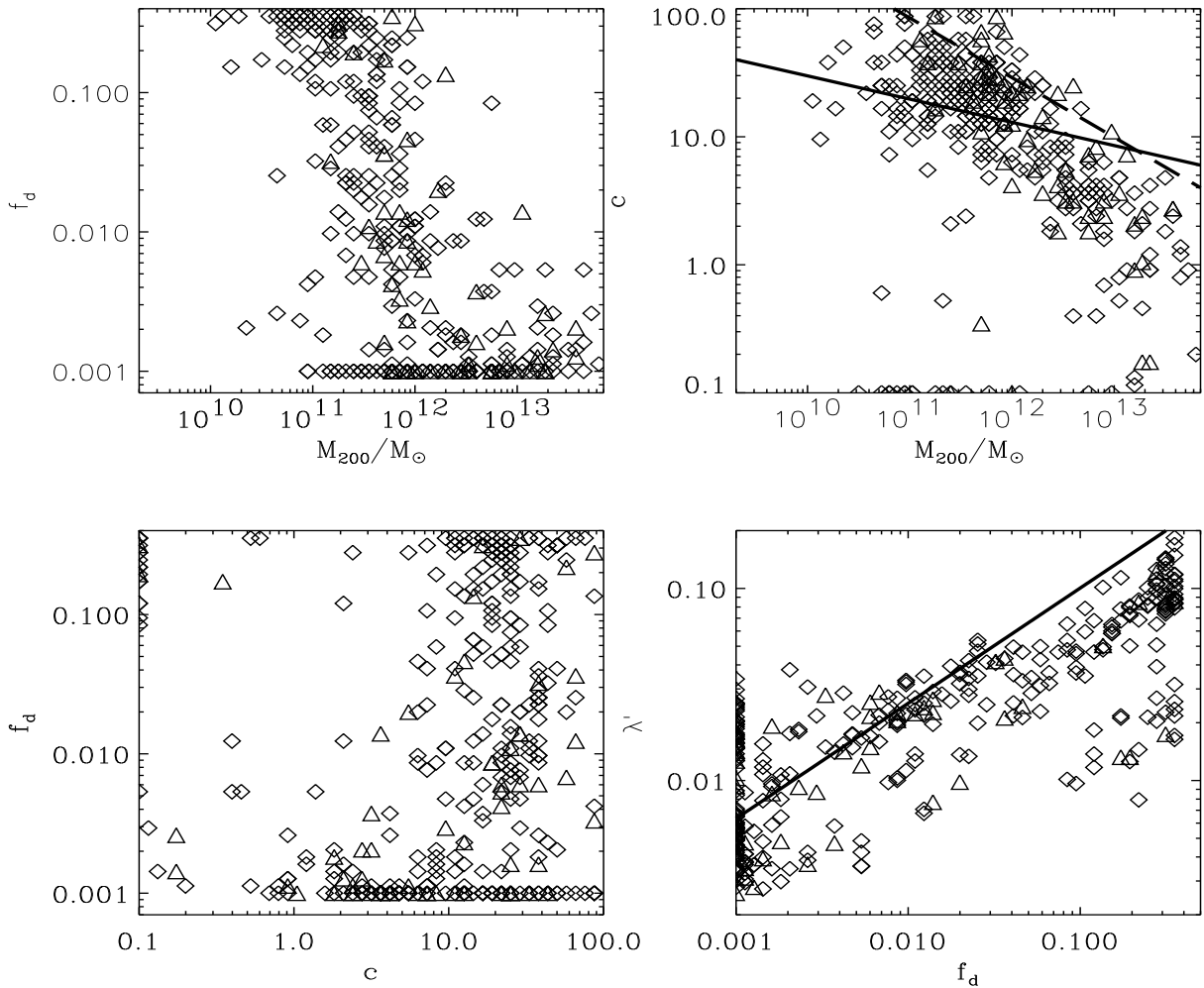}
\caption{NFW model. The location of the sample of galaxies in different
  projections of the three dimensional parameter space ($M_{200}$, $f_d$, $c$)
  when fitted with a NFW dark halo model. The diamonds correspond to galaxies
  from \citet{Courteau97} and the triangles are from \citet{PW00}.  The solid
  and dashed lines in the top-right panel are the N-body simulations
  predictions for the $c$-$M_{200}$ correlation (see text for more details).
  Also plotted (bottom-right panel) is the deduced value for $\lambda^{\prime}$
  vs.  $f_d$. The solid line is the least square fit from 14 dwarfs by
  \citet{Burkert00}, in good agreement with the correlation found here from
  362 galaxies.  Note also the correlation between $f_d$ and $M_{200}$.}
\label{fig:nfwall}
\end{figure*}

\subsection{The observational sample}

We use three sets of observed rotation curves. The first is the catalog
compiled by \citet{Courteau97} (Courteau sample) which consists of optical
$H_{\alpha}$ long--slit rotation curves for 300 Sb--Sc UGC galaxies and
$R$-band photometry. The second is a sample of 74 spiral galaxies (Sa--Sd)
observed again in the optical ($H_{\alpha}$) by \citet{PW00} (PW sample). For
this sample I--band photometry and the velocity field in two dimensions is
available. Finally, we also consider the recently published optical
($H_{\alpha}$) rotation curves for 26 Low Surface Brightness (LSB) galaxies by
\citet*{MRB01}. These consist of long slit $H_{\alpha}$ observations for 26
galaxies and $B$-band photometry for a smaller sample. In total we have 400
optical rotation curves.

All three samples provide inclination corrected curves with their
corresponding errors. The Courteau and PW samples have surface photometry and
give values for the stellar disk scale lengths, but we do not have disk scale
lengths for the LSB sample.
 
The sample of galaxies studied here is somewhat heterogeneous: we use LSB and
HSB galaxies; some curves have much smaller error-bars than others, and the
sampling range of the rotation curve varies significantly: in particular, some
curves do not extend to large enough radii to show the flattening of the
rotation curve. The diversity of the sample is both a strength as well as a
weakness: it allows us to probe general statistical trends across a wide range
of galaxy luminosities and surface brightness.

\subsection{The models}

Many authors have investigated galactosynthesis models
\citep*{DSS97,JHHP97,MMW98,JPMH98,sp99,bosch00,af00,fa00,NS00} in which the
properties of disk galaxies are determined primarily by the mass, size, and
angular momenta of the halos in which they form, and which may contain the
effects of supernova feedback, adiabatic disk contraction, cooling, merging,
and a variety of star-formation recipes.  In this paper, we are mainly
concerned with the dynamical properties of the disk/halo system which affect
the rotation curve, in the spirit of \citet{DSS97,JHHP97}, and \citet*{MMW98}
(hereafter MMW). We model spiral galaxies as exponential disks (i.e., disks
with surface density profiles $\Sigma(r)=\Sigma_0\exp(-r/R_d)$, where $R_d$ is
the scale length of the disk) embedded within CDM halos which we model either
with NFW profiles or pseudo-isothermal spheres.  In both models we assume that
the baryons initially follow the same density profile as the dark matter and
then cool and settle into an exponential disk in a dynamical time. We assume
that the baryons that do not settle in the disk follow the dark matter
distribution.  The disks have a mass fraction $f_{d} \equiv M_{\rm
  disk}/M_{200} \le \Omega_{b}/\Omega_{m}$. Here $M_{\rm disk}$ denotes the
baryonic mass that settle into an exponential disk, $M_{200}$ denotes the mass
within a radius $R_{200}$ that encloses an average density 200 times the
critical density, $\Omega_m$ denotes the density parameter and $\Omega_b$ the
baryon contribution to the density parameter.  Throughout, we assume a
$\Lambda$CDM cosmology given by:
$(\Omega_{m},\Omega_{\Lambda},\Omega_{b},h,\sigma_{8
  h^{-1}})=(0.3,0.7,0.039,0.7,1.0)$.

In these models, the rotation velocity is given by:
\begin{equation}
V_c^2(r)=V^2_d+V^2_{\rm CDM}
\label{eq.vc}
\end{equation}
where the baryonic disk rotation velocity is (\citet{BT87}, p. 77) 
\begin{equation}
V^2_d(r)=4\pi G \Sigma_0 R_d y^2 \left[I_0(y)K_0(y)-I_1(y)K_1(y)\right]\,,
\label{eq.vcdisk}
\end{equation}
with $y=r/(2R_d)$, $\Sigma_0=f_d M_{200}/(2\pi R_d^2)$, $I_n$ and $K_n$
denoting the modified Bessel functions of the first and second
kind. The contribution to the rotation curve from the dark matter is:  
\begin{equation}
V^2_{\rm CDM}(r)=GM_{\rm CDM}(r)/r,
\end{equation}
where $M(r)$ denotes the mass enclosed within the radius $r$.
For a model to be fully specified, we require expressions for the disk
scale length $R_{d}$ and for $M(r)$. 

The NFW fit to the density profile of dark matter found in collisionless
N-body numerical simulations is:
\begin{equation}
\rho(r)=\rho_{\rm crit}\frac{\delta_0}{(r/r_c)(1+r/r_c)^2}
\label{eq:rhocdm}
\end{equation}
where $\rho_{\rm crit}$ is the critical density ($\rho_{\rm crit} = 277.3 h^2$
M$_{\odot}$ kpc$^{-3}$) and $r_c$ is roughly the break radius at which the
slope of the profile changes from $-1$ to $-3$. The concentration parameter
$c$ is defined as $c=R_{200}/r_c$, where $R_{200}=(M_{200}/(4/3\pi 200
\rho_{\rm crit}))^{1/3}$ and $M_{200}$ is assumed to be the total mass of the
dark matter halo. The dark matter mass within an enclosed radius $r$ is:
\begin{equation}
M(r)=4 \pi \rho_{\rm crit} \delta_{0} r_{c}^{3} \left[ \frac{1}{1+ cx}
-1 + {\rm ln}(1+cx) \right], 
\label{eq:Mr_NFW}
\end{equation}
where $x=r/R_{200}$.
From equation (\ref{eq:Mr_NFW}), the relation between $\delta_{0}$ and
$c$ is: 
\begin{equation}
\delta_0=\frac{200}{3}\frac{c^3}{\ln(1+c)-c/(1+c)}.
\label{eq:delta0}
\end{equation}
Disk formation affects the dark matter profile by the adiabatic contraction it
induces in the inner regions of the dark matter halo. As in MMW, we model
this by assuming that the halo responds adiabatically to the slow
contraction of the disk, and therefore that the angular momentum of
dark matter particles is conserved. This yields the relation between
the initial radius $r_{i}$ and final radius $r$ where a dark matter particle
ends up:
\begin{equation}
G M_{f}(r) r = G M(r_{i}) r_{i}
\end{equation}
where $M(r_{i})$ is given by equation (\ref{eq:Mr_NFW}), and
$M_{f}(r)= \Sigma_{o} {\rm exp}(-r/R_{d})+ M(r_{i})(1-f_{d})$ is the
total final mass within $r$ (see MMW for details). 

Assuming angular momentum conservation, MMW developed a fitting formula for
$R_{d}(c,M_{200},\lambda^{\prime},f_{d})$ which we
use (see their equations (28) and (32)). Note that we use the modified
spin parameter $\lambda^{\prime} \equiv (j_{d}/f_{d}) \lambda$, where
$j_{d} \equiv J_{d}/J$ and $J_{d},J$ are the total angular momentum of
the disk and halo respectively. If the specific angular momentum of
the disk is equal to that of the halo, then $j_{d}=f_{d}$ and
$\lambda^{\prime}=\lambda$.  

In the NFW model, for a given $M_{200}$, the overdensity $\delta_{0}$ and thus
the concentration parameter $c$ depends on the collapse redshift. There is
therefore a correlation between $c$ and $M_{200}$, since more massive halos
collapse at lower redshift than less massive ones.  However, there is a fairly
wide scatter in the distribution of $c$ for fixed $M_{200}$ (e.g.,
\citet{Jing00}). Furthermore, there is growing observational evidence that the
distribution of concentration parameters might not be as predicted by CDM
(e.g., \citet{NS00,Keeton01}); one of the goals of this paper is to explore
this question further. We therefore keep $c$ as a free parameter. Thus, there
are four free parameters in our model: the spin of the dark halo $\lambda'$,
the mass of the dark halo $M_{200}$, the concentration parameter $c$ and the
fraction of the total mass in the disk $f_d$.

For the pseudo-isothermal sphere the profile of the dark matter is:
\begin{equation}
\rho=\rho_0 \left [1+ \left (\frac{r}{r_c} \right )^2 \right ]^{-1}
\label{eq:pseudo1}
\end{equation}
where $\rho_0$ denotes the finite central density.  We have developed a
similar analytic model to compute the properties of an exponential disk in
this profile (see Appendix I). In this case, we do not include the adiabatic
compression of the halo due to the baryons, since we want to use a constant
density inner core for the final dark matter distribution by fiat. In the
pseudo-isothermal profile case, the CDM rotation velocity is:
\begin{equation}
V_{\rm CDM}^2=4 \pi G \rho_0 r_c^2\left[1-\frac{r_c}{r} {\rm arctan}\left(\frac{r}{r_c}\right)\right].
\end{equation} 
A fitting formula for the disk scale length $R_{d}$ as a  function of
$M_{200}$, $c$, $f_d$ and $\lambda'$ 
for this profile is given
in Appendix I. Any rotation curve can be fitted with this profile using four
parameters: $\lambda'$, $M_{200}$, $r_{c}$, $f_d$. 

For each galaxy rotation curve we explore the whole likelihood surface
and the best fitting model is then obtained using a standard $\chi^2$
minimization.

We discuss in detail the values for the disk parameters found for these two
models (NFW and pseudoisothermal) when using all four free parameters in
Appendixes II and III. In this section we eliminate one of the four free
parameters ($\lambda'$) by using the observed scale lengths of the exponential
disks $R_d$. Then the only remaining free parameter for the baryonic disk is
$f_d$. Since we do not have $R_d$ for the LSB sample and for some of the
galaxies in the PW sample, the analysis with three free parameters is
performed on the Courteau and PW samples (in total 362 galaxies).  The disk
parameter distributions for the four parameters fit is quite similar to that
for the three parameter fit. We discuss the locus of LSB galaxies in the
parameter space in Appendix II and show that they follow the same trends as
the Courteau and PW samples.

\begin{figure}
\includegraphics[width=8.5cm,height=6cm]{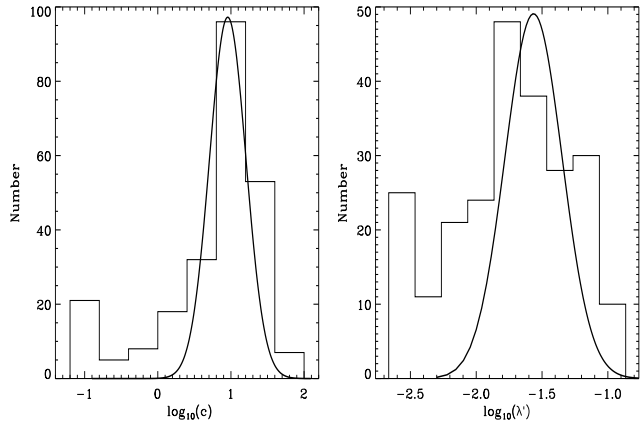}
\caption{NFW model. The distribution of $c$ (left panel) 
  and $\lambda^{\prime}$ (right panel) deduced from NFW halos fits to rotation
  curves. Also over-plotted as continuous lines are the theoretical
  predictions for each of these parameters.  Note that the derived values of
  $c$ are broadly in agreement with the predictions from N-body simulations
  (see text for more details).}
\label{fig:clamrd}
\end{figure}

\section{Best fit models}
\subsection{Fits to disk models within NFW profiles}

\begin{figure*}
\includegraphics[width=16cm,height=14cm]{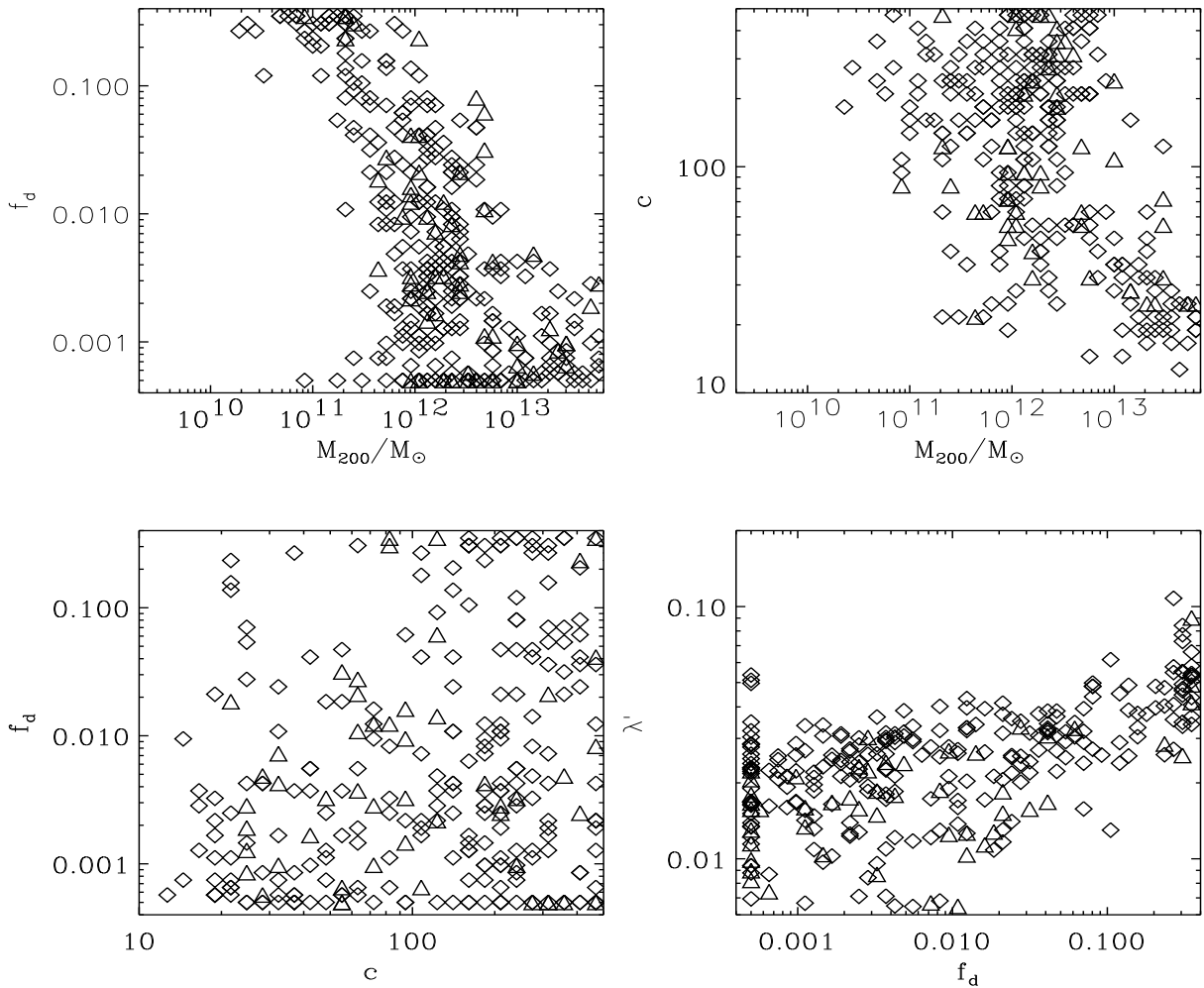}
\caption{Pseudo-isothermal model. Same as fig.~\ref{fig:nfwall} but when using
  the pseudo-isothermal dark matter halo. Note the similarity with
  fig.~\ref{fig:nfwall}.  }
\label{fig:nfwall_pseu}
\end{figure*}

Since we use the observed (stellar) scale lengths of the baryonic disk
($R_d$), the free parameters in our model are: $M_{200}$, $f_d$ and $c$.  For
each galaxy rotation curve we explore the entire likelihood surface within the
following boundaries: $10^9 <$ M$_{200}$/M$_{\odot} < 7 \times 10^{13}$, $0.1
< c < 100$ and $0.001 < f_d < 0.36$. The choice of the boundaries is justified
as follows: for $M_{200}$ by the typical velocities at large radii in the
sample, for $c$ by theoretical ansatzs given by N--body simulations and
analytic predictions (e.g., MMW); and finally for $f_d$ a realistic upper
bound can be given by nucleosynthesis constraints on the baryon fraction
$f_{d} \sim 0.2$ (e.g. \citet{Jaffe+01,Efstathiou+2dF01}); a realistic lower
bound is more difficult to set, but it is reasonable to believe that at least
few percent of the baryons in the halo cool to form the disk.  We set a higher
value $f_{d}=0.36$ and a lower value of $f_{d}=0.001$ to explore parameter
space more fully. The best fitting model is then obtained by using a standard
inverse variance weighted $\chi^2$ minimization\footnote{The best fitting
  model to all 400 rotation curves can be found at {\tt
    http://www.physics.rutgers.edu/$\sim$raulj/rot/roti.ps (rotipseu.ps)}}. We
compute the value of $\chi^{2}$ on the entire grid of points to ensure that we
converge on a global, rather than local, minimum in $\chi^{2}$. The
discreteness of points in our solution reflects the finite spacing of our
grid.

Figure~\ref{fig:nfwall} shows the position of each galaxy for all the
projections of the three dimensional parameter space
 ($M_{200}$, $f_d$ and $c$). In addition, for the best fitting values of
$M_{200}$, $f_d$ and $c$ we have computed $\lambda'$ and plot it against
$f_d$. 

For 122 galaxies out of the 362 the $\chi^2$ minimization procedure does not
find a global minimum within the boundaries for the disk parameters.  This
means that for $\sim 1/3$ (i.e., 34\%) of the galaxies in the sample the model
with a NFW profile for the dark matter fails to reproduce the observed
rotation curve. This is in agreement with a previous study by \citet*{vanBS01}.
They analyze (with a different procedure and different modelling assumptions
from ours) 20 rotation curves of late type dwarfs and find that for 6 of these
galaxies (i.e., 30\%) no meaningful fit can be obtained.  It is important to
note that our analysis uses a different approach, a much larger sample of
galaxies that is not limited to dwarfs, and still concludes that the same
percentage of galaxies are well fitted by a CDM profile.

There are some correlations among disk parameters worth commenting on. The
top-left panel of fig.~ \ref{fig:nfwall} shows an anti-correlation between
$M_{200}$ and $f_d$, with a clear zone of avoidance for low masses and low
spins (we discuss the significance and possible physical origin of this zone
of avoidance in \citet*{VOJ02}).

The top-right panel of fig.~\ref{fig:nfwall} shows that low mass halos have on
average higher concentration parameters than massive halos, in agreement with
theoretical predictions. This is not surprising since in hierarchical CDM
models low mass halos form at higher redshift than massive halos. The solid
and dashed lines in the top right panel of fig.~\ref{fig:nfwall} show the
median $M-c$ relation predicted from numerical simulations (\citet{BKSSKKPD01}
figs. 4 and 5) for low density and high density environments respectively.
Although the recovered distribution of $c$ vs. $M_{200}$ seems to show a
somewhat steeper decline of $c$ with increasing $M_{200}$ (in agreement with
the findings of \citet*{vanBBS01,vanBS01} for 14 dwarf galaxies), it is
difficult to draw a definitive conclusion due to the scatter in the theory
prediction (simulations show a 30\% scatter around this median relation) and
in the recovered correlation.  The overall distribution of halo concentrations
is in reasonable agreement with $\Lambda$CDM predictions.  This is more
clearly seen in the left panel of fig.~\ref{fig:clamrd} where the theoretical
predictions by \citet{Jing00} are overplotted as a thick line on the
distribution of $c$ (obtained from those galaxies for which a meaningful fit
was found).  Maybe more interesting is the fact that very few galaxies are
found to have values of $c$ below the theoretical predictions.  The small tail
of 20 galaxies with $c < 0.3$ is mostly due to 14 galaxies for which $f_d >
0.2$, i.e., above the value allowed by nucleosynthesis. For these galaxies the
best fit parameter combination is thus unphysical. In \S 4 we will argue that
this might be due to limitations of our model for the rotation curves.

The bottom-left panel shows that there is no evidence for a correlation
between $f_d$ and $c$. On the other hand (bottom-right panel) there is a clear
correlation between $\lambda^{\prime}$ and $f_d$. This is in agreement
with that found by \citet{Burkert00} (the solid line shows his best least
square fit to the data) and \citet*{vanBBS01} but for
only 14 dwarf galaxies. This is puzzling because cosmological models do not
predict any  correlation between $\lambda$ (where $\lambda'=j_d/f_d\lambda$)
and $f_d$.

\citet{Burkert00} points out that this correlation can arise because of a
correlation between the specific disk angular momentum ($j_d/f_d$) and $f_d$.
If baryons that make up the disk cooled from the inside out, the outermost
rings of the disk, that contain most angular momentum, cooled later; thus
predicting a correlation with a slope very close to that observed.
In this scenario the total specific angular momentum of the disk should be
significantly smaller than that of the dark matter, since most of the galaxies
have $f_d$ below nucleosynthesis value, but this seems not to be the
case for the 14 galaxies used in the analysis \citep*{vanBBS01,Burkert00}.

On the right panel of fig.~\ref{fig:clamrd_pseu} we show the distribution of
$\lambda^{\prime}$ for galaxies of our larger sample  for which a meaningful
fit was found.  Also plotted as a solid thick line is the theoretical
prediction \citep{BDKKKPP01} for the distribution of $\lambda$ found in N-body
simulations (which is very similar to that predicted by linear theory
\citep{HP88}). Although the agreement is reasonable, there is a 
tail of galaxies for which the recovered $\lambda^{\prime}$ is below the
predictions.

\citet{Burkert00} did not find this, but this may be due to the fact that the
galaxies in the  sample of \citet*{vanBBS01} are LSB's so they are selected to
have high spin \citep{JPMH98}.

We argue that the most significant feature of the bottom-right panel of figure
1 is the lack of galaxies above the solid line (what we call zone of
avoidance). Many different effects might be at play here: disks in the bottom
right portion of the plot may be unstable to bulge formation and become
bulge-dominated; disks may form preferentially out of low angular momentum
gas, which settles in the center first; finally for some combinations of the
parameters disks might fail to trigger star formation.  In \citet*{VOJ02} we
discuss in more detail the origin of the zone of avoidance.

In Appendix II, we explore the degeneracies between model parameters.
Although we find that there should be no significant degeneracies between
model parameters for most of these high-quality rotation curves, we find that
for the lowest quality rotation curves, significant degeneracies develop. The
degeneracies have almost exactly the same direction as the 3 correlations
recovered from the data. We therefore caution the reader that the correlations
between model parameters seen above maybe the result of parameter degeneracies
if there are systematic errors in the rotation curves or the quoted errors
have been underestimated.

\subsection{Fits to disk models within a Pseudoisothermal halo}

We fit the pseudoisothermal halo model of \S 2.1 to the same set of
galaxies as above using the same inverse variance weighted $\chi^2$. Our aim
is two-fold: first to see if the same correlations among the parameters found
for the NFW profile still hold for this new profile and second, to investigate
the distribution of core sizes.

We find that, for this dark matter profile model, for 114 galaxies out of the
362 (i.e. 32\%) the $\chi^2$ minimization procedure does not find a global
minimum within the boundaries.

Figure~\ref{fig:nfwall_pseu} shows the location of galaxies for the
projections of the three dimensional parameter space ($M_{200}$, $f_d$, $c$),
where $c \equiv R_{200}/r_{c}$. As for the NFW profile, there is a clear
anti-correlation between $f_d$ and $M_{200}$ and the same zone of avoidance.
In \citet*{VOJ02} we discuss in detail the physical origin of this zone of
avoidance. Here we only point out that this zone of avoidance seems to be
independent of the shape of the dark matter profile.

\begin{figure}
\includegraphics[width=8.5cm,height=6cm]{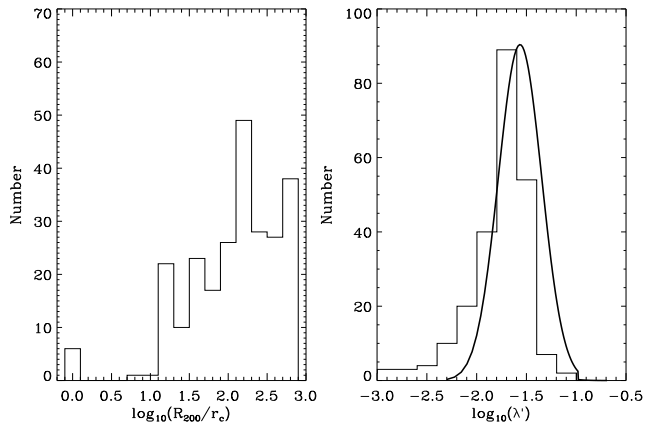}
\caption{Pseudo-isothermal model. Left panel: distribution of the 
  concentration parameter $c=R_{200}/r_c$ for the pseudo-isothermal dark halo.
  Note the sharp cut-off at $c=16$, 70\% galaxies in the sample have $c >
  100$, i.e.  70\% of galaxies have cores with sizes below 1\% of the dark
  halo virial radius.  Right panel: the distribution of $\lambda^{\prime}$.
  The continuous line is the numerical simulations prediction (see text for
  more details).}
\label{fig:clamrd_pseu}
\end{figure}

There is no evidence for a correlation between $f_d$ and $c$.  As for the NFW
profile we can now use the best fitting values of ($M_{200}$, $f_d$ and $c$)
to compute the value of $\lambda^{\prime}$ using eqs. 19 and 20 from Appendix
I. There is a clear correlation between $f_d$ and $\lambda'$ (bottom right
panel of fig.~\ref{fig:nfwall_pseu}), very similar to that found for the NFW
model (fig.~\ref{fig:nfwall}).  The distribution of $\lambda'$ for the whole
sample is shown on the right panel of fig.~\ref{fig:clamrd_pseu}; the thick
solid line is the theoretically predicted value. Although the theoretical
prediction is obtained from numerical simulations, where the dark matter halo
profile is closer to the NFW, there is reasonable agreement. This is not
surprising since the N-body predictions are very similar to the linear theory
ones, and if halos were isolated (i.e., no further merging after formation)
they would be well approximated by isothermal spheres (e.g., \citet{GG72}).

The upper-right panel shows that most galaxies in the sample have
$c=R_{200}/r_c$ larger than 20, already pointing to small sizes of the cores.
This is better illustrated in the left panel of fig.~\ref{fig:clamrd_pseu},
where we show the histogram of $R_{200}/r_c$ for the galaxies for which
meaningful fit are obtained.  There is a sharp cut--off at $R_{200}/r_c \sim
16$, with 70\% of the galaxies having $R_{200}/r_c > 100$. This implies that
the size of dark matter cores is modest, about 6\% of the virial radius or
smaller.

\begin{figure}
\includegraphics[width=8.5cm,height=8cm]{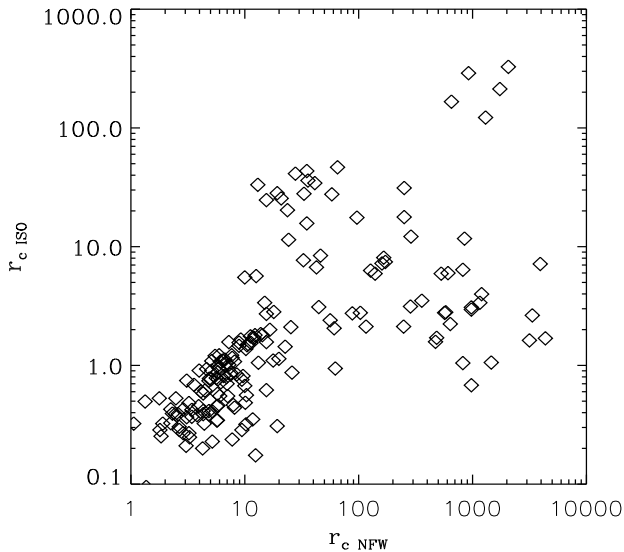}
\caption{The size of the core radius in the pseudo-isothermal model compared to
  the break radius in the NFW model. Note that the former is 10 times smaller
  than the latter and that 70\% of the sample has cores below 2 kpc.}
\label{fig:rcvsrc}
\end{figure}

\section{General considerations}

From the recovered best fitting disk parameters in the context of the 2 models
(NFW profile and pseudo-isothermal profile) we can draw some general
conclusions.  First of all, the correlations between $c$ and $M_{200}$, $f_d$
and $M_{200}$ and $f_d$ and $\lambda'$ are quite similar for the two dark
matter profiles: these correlations and the zone of avoidance in the different
projections of the three dimensional parameter space ($M_{200}$, $c$, $f_d$)
seem to be independent from the shape of the dark matter profile.  The disk
parameter correlations we find from this large (362 galaxies) and diverse
sample of galaxies are in agreement with those found from previous studies
that involved much smaller samples ($\sim 20$) and only dwarf or LSBs galaxies
\citep*{Burkert00,vanBBS01}.

Both NFW and pseudoisothermal profiles are {\em good fits to about 70\% the
  rotation curves}.  For the NFW profile we find no meaningful fit for 34\% of
the galaxies (122 out of 362), while for the pseudoisothermal profile we find
no meaningful fit for 32\% (114 out of 362).  We do not expect to find that
the model provides a good fit for all the galaxies in the sample: galaxies
might be more complicated than our model for the rotation curves. In fact some
rotation curves show strong evidence for spiral arms and bars in the rotation
pattern, and we have not attempted to model any bulge or bar component.
However, we find that many rotation curves that have no meaningful fit for one
dark matter profile are well fitted by the other: for only 52 galaxies (14\%
of the sample) there is no meaningful fit in either model. These correspond to
the poorest quality rotation curves (low spatial resolution and poor spatial
coverage, i.e. no flat or inner rotation curve).

Fig.~\ref{fig:rcvsrc} shows that the core radius in the pseudo-isothermal
profile ($r_{c \,{\rm ISO}}$) is much smaller than the break radius in the NFW
profile ($r_{c \,{\rm NFW}}$): $r_{c \,{\rm ISO}}$ is about $10$ times smaller
than $r_{c \,{\rm NFW}}$.  In other words, best fitting models for the dark
matter in this sample of galaxies are those with steep dark matter profiles
($r^{-1}$ or steeper), at least down to 6\% of the virial radius.

\begin{figure}
\includegraphics[width=8.5cm,height=6cm]{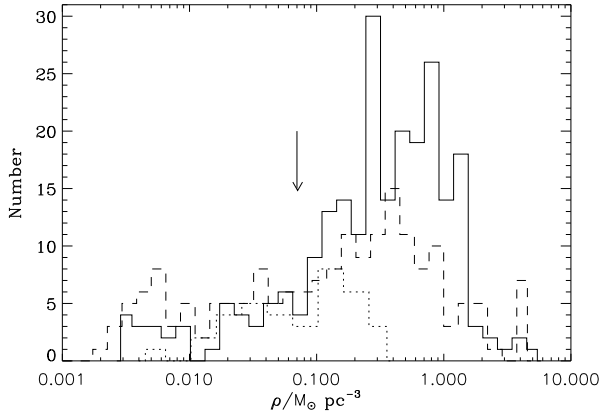}
\caption{The distribution of central  dark matter density for the two 
  profiles at a distance from the center of 0.5 kpc. The dashed line
  corresponds to pseudo-isothermal dark halos while the solid line corresponds
  to NFW dark halos. Note the modest density of dark matter. The arrow shows
  the central density (at $r=0.5$) inferred by \citet{BS01} and the dotted
  line is the distribution of central densities for 27 galaxies from
  \citet{BMBR01} that show a flat core.}
\label{fig:densities}
\end{figure}

The above considerations lead us to conclude that large dark matter cores (and
therefore low central dark matter densities) seem to be excluded.  In
fig.~\ref{fig:densities} we show the distribution of central dark matter
densities recovered from the best fits to the rotation curves.  For the NFW
profile model, since the central density is not finite, the densities have
been evaluated at $r=0.5$ kpc. This is comparable to the smallest radius at
which the rotation curve can be reliably measured in the samples used in this
papers.

The arrow shows the central density (at $r=0.5$ kpc) inferred by
\citet{BS01} (see their fig. 5) from 9 galaxies.  

We have also selected 37 galaxies (out of the 48 LSBs) analyzed by 
\citet{BMBR01}, that show indication of a flat core, and computed the
distribution of inferred central densities (the 9 galaxies which show
no flattening of the profile should have higher central densities). This
distribution is shown as the dotted line in fig.~\ref{fig:densities}.

There are a few features worth commenting on. First, the mean density for the
pseudo-isothermal profile is slightly lower than that for the NFW profile at
0.5 kpc. The average value of derived central densities for the 27 core
galaxies selected from \citet{BMBR01} is not significantly lower than the
average value for the pseudo-isothermal profile. Therefore, there is quite a
reasonable agreement for the central densities derived from our model and
those derived inverting the Poisson equation in the above LSB galaxies. Of
course, the NFW densities will be much higher at radii below 0.5 kpc.  Since
densities for smaller radii have not been directly derived in a reliable way,
it is difficult to assess whether or not the high dark matter densities
predicted by the NFW profile are excluded. We simply point out that physical
mechanisms that remove efficiently most of the dark matter inside the central
kpc from a steep dark matter profile are currently under investigation
(Merritt et al. 2002, in preparation; Milosavljevic et al. 2002, in
preparation).

\subsection{$M/L$ of the sample}

A good independent test of the values of the best fit disk parameters (in
particular of $f_{d}$) recovered from modelling of the rotation curves is a
comparison of the baryonic mass-to-light ratios ($M/L$) inferred from the
recovered disk mass $M=f_{d} M_{200}$ and measured luminosity ($L$; from
photometry) with values predicted by synthetic stellar population models.
Stellar synthesis models assume an initial mass function (IMF) and a star
formation rate, age and metallicity to uniquely predict the $M/L$. The IMF is
an ansatz which can be guided by observations in the solar neighborhood
(e.g., \citet{Kroupa01}), while the star formation rate, age and metallicity
can be inferred with help of the spectrum of the stellar population. In our
case, because spectra are not available, we have to make reasonable guesses
for these parameters. Once they have been fixed, current stellar synthesis
models agree, quite remarkably, in their predictions about the $M/L$ ratio
(e.g., \citet*{BJ01}). However, uncertainty in the assumptions ultimately limit
the precision in the $M/L$ predictions from synthetic stellar population
models, leaving a 20-40 \% uncertainty.  Taking these uncertainties into
account, a large disagreement between the $M/L$ recovered from the rotation
curve modelling and permitted values from stellar synthesis models, would
imply that the modelling of the rotation curves is probably incorrect.

\begin{figure}
\includegraphics[width=8.5cm,height=6cm]{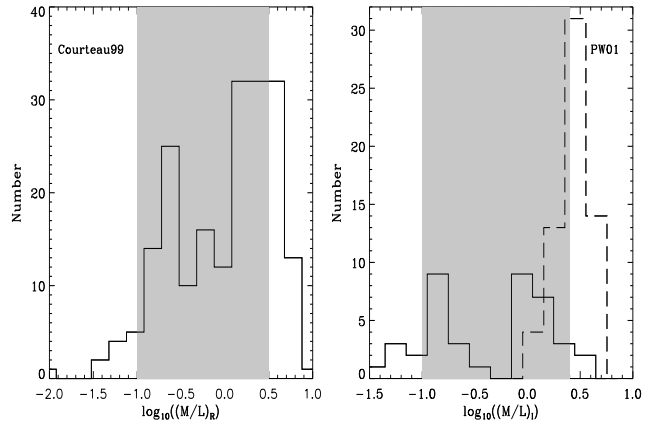}
\caption{NFW model. The predicted (M/L) in our analysis, using the NFW 
  dark halo profile, for the Courteau and PW samples. The left panel
  corresponds to the Courteau sample.  The right panel shows the histogram for
  the galaxies in the PW sample as a solid line. The dashed line shows the
  resulting distribution for the (M/L) ratio assuming a maximum disk model
  \citep{PW00}. The gray areas correspond to the allowed values from synthetic
  stellar populations.}
\label{fig:mlcdmrd}
\end{figure}

\begin{figure}
\includegraphics[width=8.5cm,height=6cm]{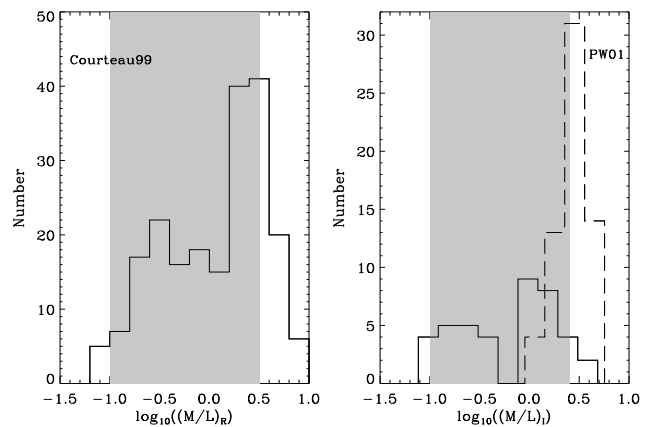}
\caption{Pseudo--isothermal model. The predicted (M/L) in our analysis and 
  in synthetic stellar populations. The line styles and shaded areas have the
  same meaning as in fig.~\ref{fig:mlcdmrd}.}
\label{fig:mlpseurd}
\end{figure}

Fig.~\ref{fig:mlcdmrd} shows the baryonic $M/L$ ratios derived by using the
best fitting model values for $M_{200}$ and $f_d$ for the NFW profile (for the
240 galaxies for which meaningful fits were found) and the published values
for the reddening-corrected integrated photometry for the Courteau and PW
samples \citep{Courteau97,PW00}; Fig~\ref{fig:mlpseurd} shows the distribution
of $M/L$ ratios recovered by using the pseudo-isothermal profile model. In
both cases, the recovered distribution of $M/L$ is fairly broad.

In both figures, the allowed range of $M/L$ using the synthetic stellar
population models developed in \citet{JPMH98} are shown as gray areas. This
range is obtained by varying $B-R$ (between 0 and 1.5), the IMF (Salpeter;
Miller--Scalo) and the star formation history of galaxies (a star formation
recipe with a star formation rate $\dot{M}_{*} \propto {\rm exp}(-t/t_{*})$,
where $3$ Gyr $\le t_{*} \le 20$Gyr). These assumptions allow $0.1< \left( M/L
\right)_{R,I} < 3$ where the subscript R, I refers to the $M/L$ in the R and I
bands; the assumed age of the stellar population has the greatest impact on
the derived $M/L$ range, with low values of $M/L$ associated with
progressively younger stellar populations.

The $M/L$ derived from the NFW profile show a somewhat broader range than
allowed: for the Courteau sample 4\% of the galaxies have uncomfortably low
$M/L$ while 16\% have $M/L$ {\it larger} than allowed from synthetic stellar
populations; for the PW sample 14\% have low $M/L$ and 7\% have too high
$M/L$.  For the pseudoisothermal profile the situation is very similar: for
the the Courteau sample 2\% of galaxies have low $M/L$ and 17\% have too high
$M/L$ while for the PW sample 5\% have too low $M/L$ and 10\% too high.

It is commonly claimed in the literature that NFW fits require $M/L$ ratios
which are too low; this is then used as an argument against dark matter
profiles with central cusps \citep*{BMR01}. The majority of studies have
however focused on LSBs; we find that for HSBs only a small fraction of
galaxies have excessively low $M/L$. Low values of $M/L$ may plausibly be
associated with very young stellar populations (e.g., if the last burst of
star formation is $\sim 1$ Gyr old, then $(M/L)_{\rm R,I} \sim 0.1$. Tighter
constraints on $M/L$ are only possible with multi-color photometry or
ideally, spectroscopy). Note that the distributions of $M/L$ ratios derived
from the pseudoisothermal profile are very similar to those derived using the
NFW profile (fig.~\ref{fig:mlcdmrd} and \ref{fig:mlpseurd}); removing the
central dark matter cusp does not change $M/L$ significantly.  Thus we
conclude that the derived disk mass is fairly independent of the dark matter
profile we assume and in broad agreement with values allowed by synthetic
stellar population models.

A possible criticism of studying HSB galaxies is that, in contrast to LSBs,
the rotation curves of these galaxies are not dark matter dominated. In an
extreme case, this would correspond to a maximum disk model, in which the
circular velocity is attributable solely to the disk. In this case, we would
be completely insensitive to the dark matter profile. The dashed line
histograms in the right panels of figures ~\ref{fig:mlcdmrd} and
\ref{fig:mlpseurd} show the $M/L$ ratio derived by \citet{PW00} fitting a
maximum disk model to their sample. Not surprisingly, their derived average
$M/L$ is larger than that of both the NFW and pseudoisothermal profile;
however, it is also larger than allowed by the synthetic stellar population
models. The Sa-Sd disk galaxies in the \citet{PW00} sample have $B-R < 1.2$,
which fairly robustly implies $(M/L)_{\rm R,I}<2.5$. The $M/L$ ratios implied
by the maximum disk models are much larger: they are more typically associated
with ellipticals.

\subsection{Low recovered disk mass fractions $f_d$}

It is worth commenting on the low values of $f_{d}$ we obtain. For both the
NFW and pseudo-isothermal fits, we often obtain values of $f_{d}$ as low as
$f_{d} \sim 10^{-3}$, even for $M_{200} \sim {\rm few} \times 10^{12}
M_{\odot}$. This is implausibly low: it implies only $\sim 1\%$ of the baryons
in the halo cool to form the disk. It is difficult to think of any feedback
mechanism (e.g., supernovae explosions) which could have such an extreme
effect in inhibiting cooling or expelling baryons from the disk (e.g.,
\citet{Efstathiou00}). A possible mechanism that can considerably reduce the
amount of gas in disk galaxies is RAM pressure stripping \citep*{QMB00}.
However this mechanism works only on galaxies orbiting in clusters. We do not
have extended information about the environment of the galaxies from our
sample, but very low values of $f_d$ are found also from analysis of LSB's
galaxies (e.g. \citet*{vanBBS01}) that are not found orbiting clusters.  Thus
the ram pressure stripping mechanism seems not to be responsible for the low
values of $f_d$ we recover. 

These very low values of $f_d$ are also responsible for the tail of low $M/L$
galaxies in figures ~\ref{fig:mlcdmrd} and ~\ref{fig:mlpseurd}.
 
At present, we regard these low values of $f_{d}$ as pointing toward some
inadequacy of the simple disk-halo model, or the assumed form of the dark
matter profile. On the other hand, the high mass end of the galaxy mass
function is dominated by elliptical galaxies and it is not inconceivable that
we are selecting only those dark halos that manage to form a disk.

Since the value of $f_d$ mainly affects the inner part of the rotation curve,
we may ask: how should the dark matter profile be modified so that the
recovered $f_d$ is more meaningful?  It is clear that dark matter density in
the center needs to be reduced.  However, as discussed above, the best fit
model for the pseudoisothermal profile has very small cores (and a steeper
--$r^{-2}$-- slope outside).

We are investigating possible physical mechanisms within the CDM paradigm that
can achieve such result, thus decreasing substantially the dark matter content
of a NFW halo within the inner kpc (Merritt et al. 2002, in preparation;
Milosavljevic et al. 2002, in preparation).

\section{Conclusions}
We have analyzed a large set of high spatial resolution rotation curves of
galaxies with the goal of determining the model parameters for a disk embedded
within a cold dark matter (CDM) halo that we have modeled in two ways: either
with a NFW profile or a pseudo-isothermal profile.  To this aim we have
developed the expression for the rotation curve of an exponential disk
embedded in a pseudoisothermal halo (Appendix I).

Our study differs from previous work in two respects. Firstly, we use optical
rotation curves rather than HI rotation curves, so the rotation curves we
use are free from beam smearing and have smaller errors and higher spatial
resolution.  Secondly, while many studies have focused on relatively small
samples of dwarfs, we use a large sample (400 galaxies) spanning a wide range
in luminosity and surface brightness. 

We find that the NFW profile provides a good fit to 66\% of the galaxies in
the sample (in agreement with previous studies, \citet*{vanBS01}) that were
based on few (20) dwarf galaxies), with a distribution of recovered
concentration parameters and spin parameters broadly consistent with that
predicted by CDM numerical simulations. When the sample is fitted with an
isothermal profile with a core, 68\% of the galaxies are well fitted within
this model and the best fit model favors cores with small sizes (almost all
have $r_{c} < 0.06 R_{200}$ and $\sim 70\%$ have $r_{c} < 0.01 R_{200}$).
However, we find that many rotation curves that have no meaningful fit for one
dark matter profile are well fitted by the other: for only 52 galaxies (14\%
of the sample) there is no meaningful fit in either model. These largely
comprise low quality rotation curves. Our findings are consistent with the
inner dark matter profiles being steep (slope of -1 or steeper) down to radii
that are few percent of the virial radius. Large dark matter cores (and
therefore low central dark matter densities) seem thus to be excluded.

Numerical SPH simulations of disk formation within dark halos find that disks
loose a significant fraction of their angular momentum if: 1) the dark matter
profile is steep in the center and 2) the baryons settle in right after
virialization \citep{NSbis00}. On the other hand, we recover steep dark matter
profiles and no significant loss of angular momentum. There are several routes
to explain this.  Some authors have pointed out that if baryons are allowed to
settle well after virialization of the halo (e.g. \citet{SDTC01}), then there
is no significant loss of angular momentum. Another possibility to consider
is that current hydrodynamical simulations lack the resolution to simulate a
multi-phase interstellar medium and also to follow the detailed formation of
giant molecular clouds, and therefore the sites of star formation, within the
settling disk. It is therefore not clear that current simulations contain all
the physical ingredients needed to simulate the complex process of baryons
cooling down into a disk.

In both models, for the galaxies that have meaningful fit to the rotation
curves, the recovered baryonic mass to light ratios are broadly in agreement
with predictions from synthetic stellar populations.

From this large sample of galaxies we find several correlations among the best
fitting model parameters and some regions in the parameters space that are not
populated, we refer to those regions as zone of avoidance.  These correlations
an zone of avoidance are remarkably similar in both profiles and do not change
when the rotation curves are fitted by a model with three free parameters or
with four free parameters.  In particular we find a strong correlation between
the disk mass fraction and the spin parameter for different kind of galaxies
(HSBs and LSBs), in good agreement with findings of previous studies that
involved only few dwarf LSBs galaxies. These correlations may be a result of
parameter degeneracies if errors in the rotation curve have been
underestimated; otherwise, they could have a physical origin.

When fitting the sample with the NFW profile we obtain that for the 66\% of
galaxies for which meaningful fits to the rotation curves are found, the
distribution of concentrartion parameters is in good agreement with
predictions from N-body simulations. Also, the distribution of disk spin
parameters is in broad agreement with the distribution of spin parameters for
the dark matter predicted from N-body simulations and linear theory, but with
a small tail at low spins.

The primary motivation in previous work of studying the rotation curves of
dwarfs and/or LSB's was that they would be dark matter dominated. Although it
is quite plausible, this claim must be taken with caution.  In fact, our
analysis of HSB and more luminous galaxies, as we have seen, yields values for
the baryonic disk mass fraction very similar to previous analysis of dwarfs; a
naive interpretation of rotation curve fits alone would lead us to claim these
galaxies as dark-matter dominated as well. In Appendix II, we analyze the LSB
sample of \citet*{MRB01} and find that a somewhat larger fraction of them
cannot be fit by NFW profiles, $\sim 57 \%$ as opposed to $\sim 33\%$ for the
entire sample (those which do fit have somewhat low but theoretically
plausible concentration parameters $c$). This difference may be a 1 $\sigma$
statistical fluctuation (there are only 26 LSBs in the sample), it may reflect
the greater fidelity of LSB rotation curves to the true underlying dark matter
distribution, or it may perhaps be due to a systematic difference in the dark
matter profile of halos which undergo greater tidal torquing. On the other
hand, for the pseudoisothermal model, only one LSB has $f_d > 0.2$. This seems
to indicate that LSBs are indeed fitted better with a dark matter core. For
the LSB sample, 9 galaxies have dark matter cores with sizes below 1.5 kpc and
16 below 3 kpc. This is in fair agreement with our previous findings that the
size of the dark matter core is modest. It is also in agreement with the dark
matter core size inferred from inverting the Poisson equation using the
observed rotation curve \citep{BMBR01}. Ultimately, our sample choice involves
a trade-off: we gain increased precision in the rotation curve (which breaks
parameter degeneracies) and a much larger sample encompassing a much broader
class of galaxies, but also acquire a somewhat larger uncertainty in the disk
contribution to the rotation curve.

In general, LSB galaxies tend to have higher values of $\lambda'$ for a given
$f_d$ and lower values of $c$ for a given mass than HSB galaxies (see Appendix
II). This supports the view that the dark matter profile of LSBs might reflect
an enviroment in which dark halos undergo a higher tidal torque.

If the $M/L$ of the disk was well constrained, this ambiguity can be
eliminated: it would be possible to subtract the disk contribution to the
rotation curve and recover directly the inner profile of the dark halo by
inverting the Poisson equation (e.g., \citet{BMBR01}). The M/L can be
constrained from multi-color photometry or even better, high $S/N$ spectra of
the stellar population in galaxies (extending as close as possible to the $K$
band) together with high-spatial resolution rotation curves especially of the
inner 1-2 Kpc. A more detailed treatment of this issue will be presented in a
forthcoming paper.

\section*{acknowledgments}
We thank David Spergel for insightful comments and stimulating discussions. We
also thank Povilas Palunas and Ted Williams for making available their
rotation curve sample. LV is supported in part by NASA grant NAG5-7154. SPO is
supported by NSF grant AST-0096023. LV and RJ thank the TAPIR group at Caltech
for hospitality. This research has made use of NASA's Astrophysics Data System
Abstract Service. 

\bibliographystyle{mn2e.bst} 
\bibliography{../../STY/raul}

\begin{thebibliography}{}

\bibitem[\protect\citeauthoryear{{Avila-Reese} \& {Firmani}}{{Avila-Reese} \&
  {Firmani}}{2000}]{af00}
{Avila-Reese} V.,  {Firmani} C.,  2000, Revista Mexicana de Astronomia y
  Astrofisica, 36, 23

\bibitem[\protect\citeauthoryear{{Bell} \& {de Jong}}{{Bell} \& {de
  Jong}}{2001}]{BJ01}
{Bell} E.~F.,  {de Jong} R.~S.,  2001, ApJ, 550, 212

\bibitem[\protect\citeauthoryear{{Binney} \& {Tremaine}}{{Binney} \&
  {Tremaine}}{1987}]{BT87}
{Binney} J.,  {Tremaine} S.,  1987, {Galactic dynamics}.
Princeton, NJ, Princeton University Press, 1987, 747 p.

\bibitem[\protect\citeauthoryear{{Bode}, {Ostriker} \& {Turok}}{{Bode}
  et~al.}{2001}]{BOT01}
{Bode} P.,  {Ostriker} J.~P.,    {Turok} N.,  2001, ApJ, 556, 93

\bibitem[\protect\citeauthoryear{{Borriello} \& {Salucci}}{{Borriello} \&
  {Salucci}}{2001}]{BS01}
{Borriello} A.,  {Salucci} P.,  2001, MNRAS, 323, 285

\bibitem[\protect\citeauthoryear{{Bullock}, {Dekel}, {Kolatt}, {Kravtsov},
  {Klypin}, {Porciani} \& {Primack}}{{Bullock} et~al.}{2001}]{BDKKKPP01}
{Bullock} J.~S.,  {Dekel} A.,  {Kolatt} T.~S.,  {Kravtsov} A.~V.,  {Klypin}
  A.~A.,  {Porciani} C.,    {Primack} J.~R.,  2001, ApJ, 555, 240

\bibitem[\protect\citeauthoryear{{Bullock}, {Kolatt}, {Sigad}, {Somerville},
  {Kravtsov}, {Klypin}, {Primack} \& {Dekel}}{{Bullock}
  et~al.}{2001}]{BKSSKKPD01}
{Bullock} J.~S.,  {Kolatt} T.~S.,  {Sigad} Y.,  {Somerville} R.~S.,  {Kravtsov}
  A.~V.,  {Klypin} A.~A.,  {Primack} J.~R.,    {Dekel} A.,  2001, MNRAS, 321,
  559

\bibitem[\protect\citeauthoryear{{Burkert}}{{Burkert}}{2000}]{Burkert00}
{Burkert} A.,  2000, astro-ph/0007047

\bibitem[\protect\citeauthoryear{{Cen}}{{Cen}}{2001}]{Cen01}
{Cen} R.,  2001, ApJL, 546, L77

\bibitem[\protect\citeauthoryear{{Colombi}, {Dodelson} \& {Widrow}}{{Colombi}
  et~al.}{1996}]{CDW96}
{Colombi} S.,  {Dodelson} S.,    {Widrow} L.~M.,  1996, ApJ, 458, 1

\bibitem[\protect\citeauthoryear{{Courteau}}{{Courteau}}{1997}]{Courteau97}
{Courteau} S.,  1997, AJ, 114, 2402

\bibitem[\protect\citeauthoryear{{Dalcanton} \& {Bernstein}}{{Dalcanton} \&
  {Bernstein}}{2000}]{DB00}
{Dalcanton} J.~J.,  {Bernstein} R.~A.,  2000, in Dynamics of Galaxies: from the
  Early Universe to the Present, 15th IAP meeting held in Paris, France, July
  9-13, 1999, Eds.: Francoise Combes, Gary A. Mamon, and Vassilis Charmandaris
  ASP Conference Series, Vol. 197 p.~161

\bibitem[\protect\citeauthoryear{{Dalcanton}, {Spergel} \&
  {Summers}}{{Dalcanton} et~al.}{1997}]{DSS97}
{Dalcanton} J.~J.,  {Spergel} D.~N.,    {Summers} F.~J.,  1997, ApJ, 482, 659

\bibitem[\protect\citeauthoryear{{de Blok}, {McGaugh} \& {Rubin}}{{de Blok}
  et~al.}{2001}]{BMR01}
{de Blok} E.,  {McGaugh} S.,    {Rubin} V.,  2001, astro-ph, 0107366

\bibitem[\protect\citeauthoryear{{de Blok}, {McGaugh}, {Bosma} \& {Rubin}}{{de
  Blok} et~al.}{2001}]{BMBR01}
{de Blok} W.~J.~G.,  {McGaugh} S.~S.,  {Bosma} A.,    {Rubin} V.~C.,  2001,
  ApJL, 552, L23

\bibitem[\protect\citeauthoryear{{Efstathiou}}{{Efstathiou}}{2000}]{Efstathiou%
00}
{Efstathiou} G.,  2000, MNRAS, 317, 697

\bibitem[\protect\citeauthoryear{{Efstathiou et al.}}{{Efstathiou et
  al.}}{2001}]{Efstathiou+2dF01}
{Efstathiou et al.} G.,  2001, astro-ph/0109152

\bibitem[\protect\citeauthoryear{{Firmani} \& {Avila-Reese}}{{Firmani} \&
  {Avila-Reese}}{2000}]{fa00}
{Firmani} C.,  {Avila-Reese} V.,  2000, MNRAS, 315, 457

\bibitem[\protect\citeauthoryear{{Goodman}}{{Goodman}}{2000}]{Goodman00}
{Goodman} J.,  2000, New Astronomy, 5, 103

\bibitem[\protect\citeauthoryear{{Gunn} \& {Gott}}{{Gunn} \&
  {Gott}}{1972}]{GG72}
{Gunn} J.,  {Gott} R.,  1972, ApJ, 176, 1

\bibitem[\protect\citeauthoryear{Heavens \& Peacock}{Heavens \&
  Peacock}{1988}]{HP88}
Heavens A.~F.,  Peacock J.~A.,  1988, MNRAS, 232, 339

\bibitem[\protect\citeauthoryear{{Hu}, {Barkana} \& {Gruzinov}}{{Hu}
  et~al.}{2000}]{HBG00}
{Hu} W.,  {Barkana} R.,    {Gruzinov} A.,  2000, Physical Review Letters, 85,
  1158

\bibitem[\protect\citeauthoryear{{Jaffe et~al}}{{Jaffe et~al}}{2001}]{Jaffe+01}
{Jaffe et~al} A.~H.,  2001, Physical Review Letters, 86, 3475

\bibitem[\protect\citeauthoryear{Jimenez, Heavens, Hawkins \& Padoan}{Jimenez
  et~al.}{1997}]{JHHP97}
Jimenez R.,  Heavens A.,  Hawkins M.,    Padoan P.,  1997, MNRAS, 292, L5

\bibitem[\protect\citeauthoryear{{Jimenez}, {Padoan}, {Matteucci} \&
  {Heavens}}{{Jimenez} et~al.}{1998}]{JPMH98}
{Jimenez} R.,  {Padoan} P.,  {Matteucci} F.,    {Heavens} A.~F.,  1998, MNRAS,
  299, 123

\bibitem[\protect\citeauthoryear{{Jing}}{{Jing}}{2000}]{Jing00}
{Jing} Y.~P.,  2000, ApJ, 535, 30

\bibitem[\protect\citeauthoryear{{Kamionkowski} \& {Liddle}}{{Kamionkowski} \&
  {Liddle}}{2000}]{KL00}
{Kamionkowski} M.,  {Liddle} A.~R.,  2000, Physical Review Letters, 84, 4525

\bibitem[\protect\citeauthoryear{{Keeton}}{{Keeton}}{2001}]{Keeton01}
{Keeton} C.~R.,  2001, ApJ, 561, 46

\bibitem[\protect\citeauthoryear{{Klypin}, {Kravtsov}, {Valenzuela} \&
  {Prada}}{{Klypin} et~al.}{1999}]{KKVP99}
{Klypin} A.,  {Kravtsov} A.~V.,  {Valenzuela} O.,    {Prada} F.,  1999, ApJ,
  522, 82

\bibitem[\protect\citeauthoryear{{Kroupa}}{{Kroupa}}{2001}]{Kroupa01}
{Kroupa} P.,  2001, MNRAS, 322, 231

\bibitem[\protect\citeauthoryear{{Lahav et al.}}{{Lahav et
  al.}}{2001}]{Lahav+2dF01}
{Lahav et al.} O.,  2001, astro-ph/0112162

\bibitem[\protect\citeauthoryear{{McGaugh}, {Rubin} \& {de Blok}}{{McGaugh}
  et~al.}{2001}]{MRB01}
{McGaugh} S.,  {Rubin} V.,    {de Blok} E.,  2001, astro-ph, 0107326

\bibitem[\protect\citeauthoryear{{McGaugh} \& {de Blok}}{{McGaugh} \& {de
  Blok}}{1998}]{MB98}
{McGaugh} S.~S.,  {de Blok} W.~J.~G.,  1998, ApJ, 499, 41

\bibitem[\protect\citeauthoryear{{Mo}, {Mao} \& {White}}{{Mo}
  et~al.}{1998}]{MMW98}
{Mo} H.~J.,  {Mao} S.,    {White} S. D.~M.,  1998, MNRAS, 295, 319

\bibitem[\protect\citeauthoryear{{Moore}, {Ghigna}, {Governato}, {Lake},
  {Quinn}, {Stadel} \& {Tozzi}}{{Moore} et~al.}{1999}]{MGGLQST99}
{Moore} B.,  {Ghigna} S.,  {Governato} F.,  {Lake} G.,  {Quinn} T.,  {Stadel}
  J.,    {Tozzi} P.,  1999, ApJL, 524, L19

\bibitem[\protect\citeauthoryear{{Navarro}}{{Navarro}}{1998}]{Navarro98}
{Navarro} J.,  1998, astro-ph/9807084

\bibitem[\protect\citeauthoryear{{Navarro}, {Frenk} \& {White}}{{Navarro}
  et~al.}{1997}]{NFW97}
{Navarro} J.~F.,  {Frenk} C.~S.,    {White} S. D.~M.,  1997, ApJ, 490, 493

\bibitem[\protect\citeauthoryear{{Navarro} \& {Steinmetz}}{{Navarro} \&
  {Steinmetz}}{2000a}]{NSbis00}
{Navarro} J.~F.,  {Steinmetz} M.,  2000a, ApJ, 538, 477

\bibitem[\protect\citeauthoryear{{Navarro} \& {Steinmetz}}{{Navarro} \&
  {Steinmetz}}{2000b}]{NS00}
{Navarro} J.~F.,  {Steinmetz} M.,  2000b, ApJ, 528, 607

\bibitem[\protect\citeauthoryear{{Palunas} \& {Williams}}{{Palunas} \&
  {Williams}}{2000}]{PW00}
{Palunas} P.,  {Williams} T.~B.,  2000, AJ, 120, 2884

\bibitem[\protect\citeauthoryear{{Peacock et~al}}{{Peacock
  et~al}}{2001}]{Peacock+01}
{Peacock et~al} J.~A.,  2001, Nature, 410, 169

\bibitem[\protect\citeauthoryear{{Phillips}, {Weinberg}, {Croft}, {Hernquist},
  {Katz} \& {Pettini}}{{Phillips} et~al.}{2001}]{Croft+01}
{Phillips} J.,  {Weinberg} D.~H.,  {Croft} R.~A.~C.,  {Hernquist} L.,  {Katz}
  N.,    {Pettini} M.,  2001, ApJ, 560, 15

\bibitem[\protect\citeauthoryear{{Quilis}, {Moore} \& {Bower}}{{Quilis}
  et~al.}{2000}]{QMB00}
{Quilis} V.,  {Moore} B.,    {Bower} R.,  2000, Science, Volume 288, Issue
  5471, pp.~1617-1620 (2000)., 288, 1617

\bibitem[\protect\citeauthoryear{{S{\' a}iz}, {Dom{\' i}nguez-Tenreiro},
  {Tissera} \& {Courteau}}{{S{\' a}iz} et~al.}{2001}]{SDTC01}
{S{\' a}iz} A.,  {Dom{\' i}nguez-Tenreiro} R.,  {Tissera} P.~B.,    {Courteau}
  S.,  2001, MNRAS, 325, 119

\bibitem[\protect\citeauthoryear{{Salucci}}{{Salucci}}{2001}]{Salucci01}
{Salucci} P.,  2001, MNRAS, 320, L1

\bibitem[\protect\citeauthoryear{{Salucci} \& {Burkert}}{{Salucci} \&
  {Burkert}}{2000}]{SB00}
{Salucci} P.,  {Burkert} A.,  2000, ApJL, 537, L9

\bibitem[\protect\citeauthoryear{{Somerville} \& {Primack}}{{Somerville} \&
  {Primack}}{1999}]{sp99}
{Somerville} R.~S.,  {Primack} J.~R.,  1999, MNRAS, 310, 1087

\bibitem[\protect\citeauthoryear{{Spergel} \& {Steinhardt}}{{Spergel} \&
  {Steinhardt}}{2000}]{SS00}
{Spergel} D.~N.,  {Steinhardt} P.~J.,  2000, Physical Review Letters, 84, 3760

\bibitem[\protect\citeauthoryear{{Swaters}, {Madore} \& {Trewhella}}{{Swaters}
  et~al.}{2000}]{SMT00}
{Swaters} R.~A.,  {Madore} B.~F.,    {Trewhella} M.,  2000, ApJL, 531, L107

\bibitem[\protect\citeauthoryear{{van den Bosch}}{{van den
  Bosch}}{2000}]{bosch00}
{van den Bosch} F.~C.,  2000, ApJ, 530, 177

\bibitem[\protect\citeauthoryear{{van den Bosch}, {Burkert} \& {Swaters}}{{van
  den Bosch} et~al.}{2001}]{vanBBS01}
{van den Bosch} F.~C.,  {Burkert} A.,    {Swaters} R.~A.,  2001, MNRAS, 326,
  1205

\bibitem[\protect\citeauthoryear{{van den Bosch}, {Robertson}, {Dalcanton} \&
  {de Blok}}{{van den Bosch} et~al.}{2000}]{vanBRDB00}
{van den Bosch} F.~C.,  {Robertson} B.~E.,  {Dalcanton} J.~J.,    {de Blok}
  W.~J.~G.,  2000, AJ, 119, 1579

\bibitem[\protect\citeauthoryear{{van den Bosch} \& {Swaters}}{{van den Bosch}
  \& {Swaters}}{2001}]{vanBS01}
{van den Bosch} F.~C.,  {Swaters} R.~A.,  2001, MNRAS, 325, 1017

\bibitem[\protect\citeauthoryear{{Verde}, {Oh} \& {Jimenez}}{{Verde}
  et~al.}{2002}]{VOJ02}
{Verde} L.,  {Oh} S.~P.,    {Jimenez} R.,  2002, astro-ph

\bibitem[\protect\citeauthoryear{{Verde et al.}}{{Verde et
  al.}}{2001}]{Verde+2dF01}
{Verde et al.} L.,  2001, astro-ph/0112161

\end{thebibliography}

\section*{Appendix I: Rotation curve for pseudo--isothermal profile}

In this Appendix, we derive three relations used in the text for the
pseudo--isothermal halo profile: (i) the rotation curve of the
pseudo-isothermal sphere, equation (\ref{eq:pseudoisoVcdm}), (ii) the relation
between the finite central density and the concentration parameter ($c \equiv
r_{200}/r_{c}$), equation (\ref{eq:pseudoisocentralrho}) ; (iii) a fitting
formula for the disk scale-length $R_{d}$ as a function of the halo mass and
core-radius $(M_{200}, r_{c})$, and the disk mass fraction and spin
$(f_{d},\lambda^{\prime})$ assuming angular momentum is conserved, equation
(\ref{eq:fitRd}).

Starting from 
\begin{equation}
\rho(r)=\frac{\rho_0}{1+(r/r_c)^2}
\label{eq:pseudoisoprofile}
\end{equation}
we obtain 
\begin{equation}
M(r)=\int_0^r r'^2\rho(r') dr=\rho_0 4\pi r_c^2\left[r-r_c {\rm arctan}\left(\frac{r}{r_c}\right)  \right]
\label{eq:pseudoisomofr}
\end{equation}
thus 
\begin{equation}
V^2_{CDM}(r)=4 \pi G \rho_0 r_c^2\left[1-\frac{r_c}{r}{\rm
arctan}\left(\frac{r}{r_c}\right) \right]
\label{eq:pseudoisoVcdm}
\end{equation}
and
\begin{equation}
M_{200}=\rho_0 4\pi r_c^3\left[c-{\rm arctan}(c) \right]
\label{eq:pseudoisoM200}
\end{equation}
where $c=R_{200}/r_c$.
Since by definition
\begin{equation}
\frac{M_{200}}{4/3 \pi R_{200}^3}\equiv  200 \rho_{\rm crit}
\end{equation}
we have that
\begin{equation}
\rho_0=\frac{200}{3}\rho_{\rm crit}\frac{c^3}{c-{\rm arctan}(c)}
\label{eq:pseudoisocentralrho}.
\end{equation}

The disk scale length $R_{d}$ is obtained as follows. From the
definition of the spin parameter
$\lambda$, the total angular momentum of the dark matter halo is:
\begin{equation}
J =\frac{\lambda}{|E|^{1/2}G^{-1}M_{200}^{-5/2}},
\label{eq:Jhalo}
\end{equation}
where $E$ is the total energy of the halo. The total angular
momentum of the disk $J_d$ is given by: 
\begin{equation}
J_d=2\pi\int_0^{r_{\infty}}V(r)\Sigma(r)r^2dr \equiv j_{d} J
\end{equation}
where $\Sigma(r)=\Sigma_0\exp[-r/R_d]$, and the central density is
$\Sigma_{o}=f_d M_{200}/(2\pi R_d^2)$. The total energy $E$ can be obtained
through the  virial theorem $2K+U=0$, where $k$ denotes the  kinetic energy
$K$ and $U$ the potential energy: $E=k+U=U/2$, thus,
\begin{eqnarray}
E & = & -\frac{1}{8 \pi G} \int_0^{R_{200}} \left( \frac{G M(r)}{r^{2}} \right)^{2} 4 \pi r^2 dr\\ \nonumber
& = & -\rho_0^2\pi^2 8 G r_c^4 \int_{0}^{R_{200}} dr
\left[1-\frac{r_{c}}{r}{\rm arctan} \left(\frac{r}{r_{c}}\right) \right]^{2}
\label{eq:pseudoisoEnergy}
\end{eqnarray}
Using equations (\ref{eq:Jhalo})--(\ref{eq:pseudoisoEnergy}) we obtain the following equation:
\be
q \equiv \frac{\lambda' G^{1/2}M_{200}^{3/2}}{r_c^2\sqrt{|E|4 \pi
\rho_0}}=Y^2\!\!\int_{0}^{\infty}\!\!\!\sqrt{1-\frac{{\rm
arctan}(x)}{x}}x^2\exp(-xY)dx
\label{eq:pseudoisonumRd}
\ee
where $\lambda'=\lambda j_d/f_d$ and $Y=r_c/R_d$.
Equation (\ref{eq:pseudoisonumRd}) can easily be solved numerically for $Y$ and
thus $R_d$.
We give here a fitting formula that is good to $\lap 5$\%:
\be
Y=\frac{\alpha}{q^{\beta}+{\gamma}q^{\delta}}
\label{eq:fitRd}
\ee
where $q$ is defined in eq. \ref{eq:pseudoisonumRd}, $\alpha=2.3$,
$\beta=0.4745$, $\gamma=0.85$, $\delta=1.045$. Equation
(\ref{eq:fitRd}) gives an expression for the disk scale-length $R_{d}$
for a disk embedded in a pseudo-isothermal sphere. 

\begin{figure*}
\includegraphics[width=17cm,height=20cm]{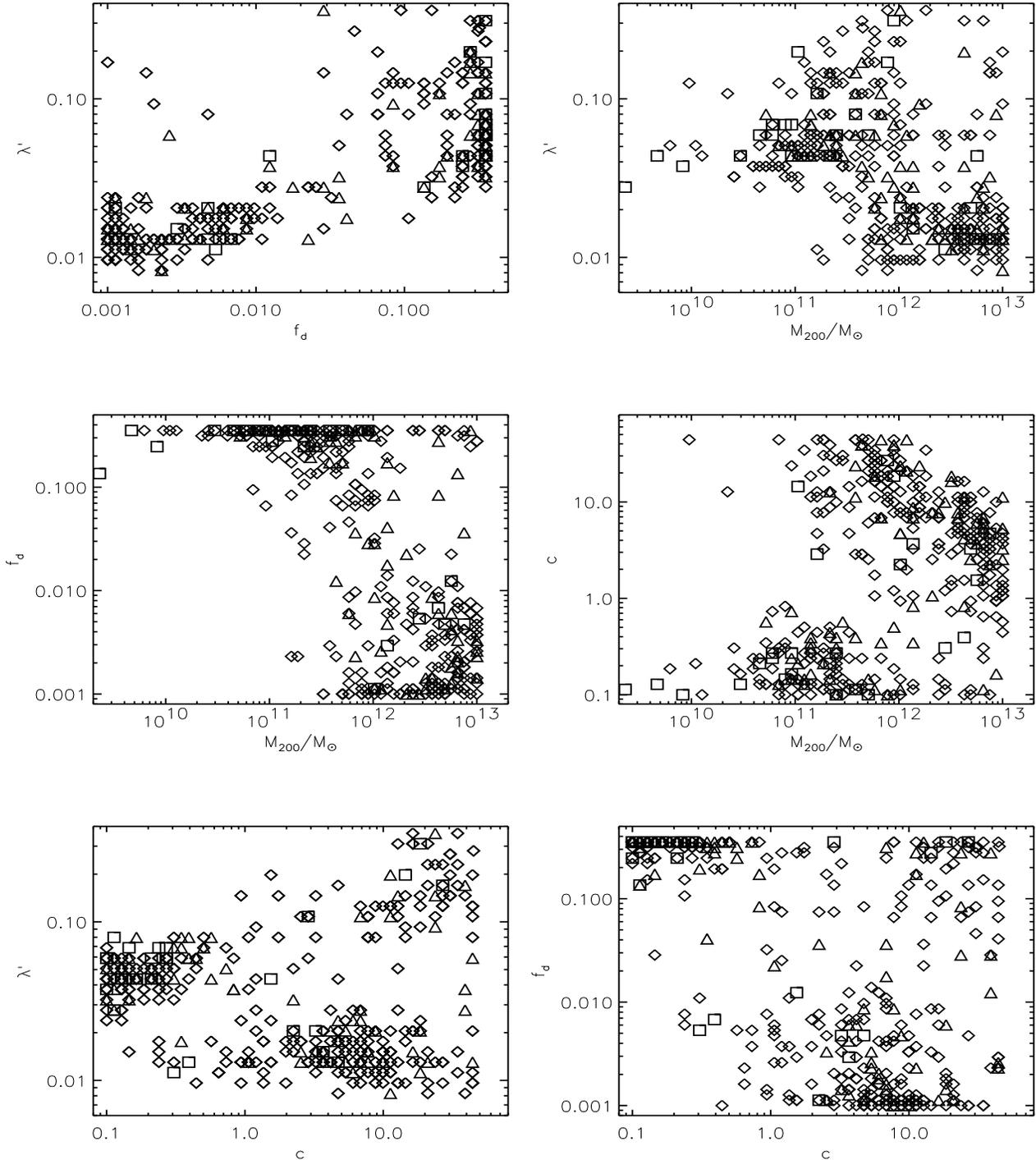}
\caption{NFW model, four parameters fit. In all six panels
  diamonds correspond to the Courteau, triangles to the PW and squares to the
  LSB sample.  The correlations between $\lambda$ and $f_d$ (upper-left
  panel), $M_{200}$ and $f_d$ (middle-left panel) are very similar to that
  obtained for the three parameter fit. Note the distribution of $c$ vs.
  $M_{200}$.}
\label{fig:hist}
\end{figure*}

\begin{figure*}
\includegraphics[width=17cm,height=20cm]{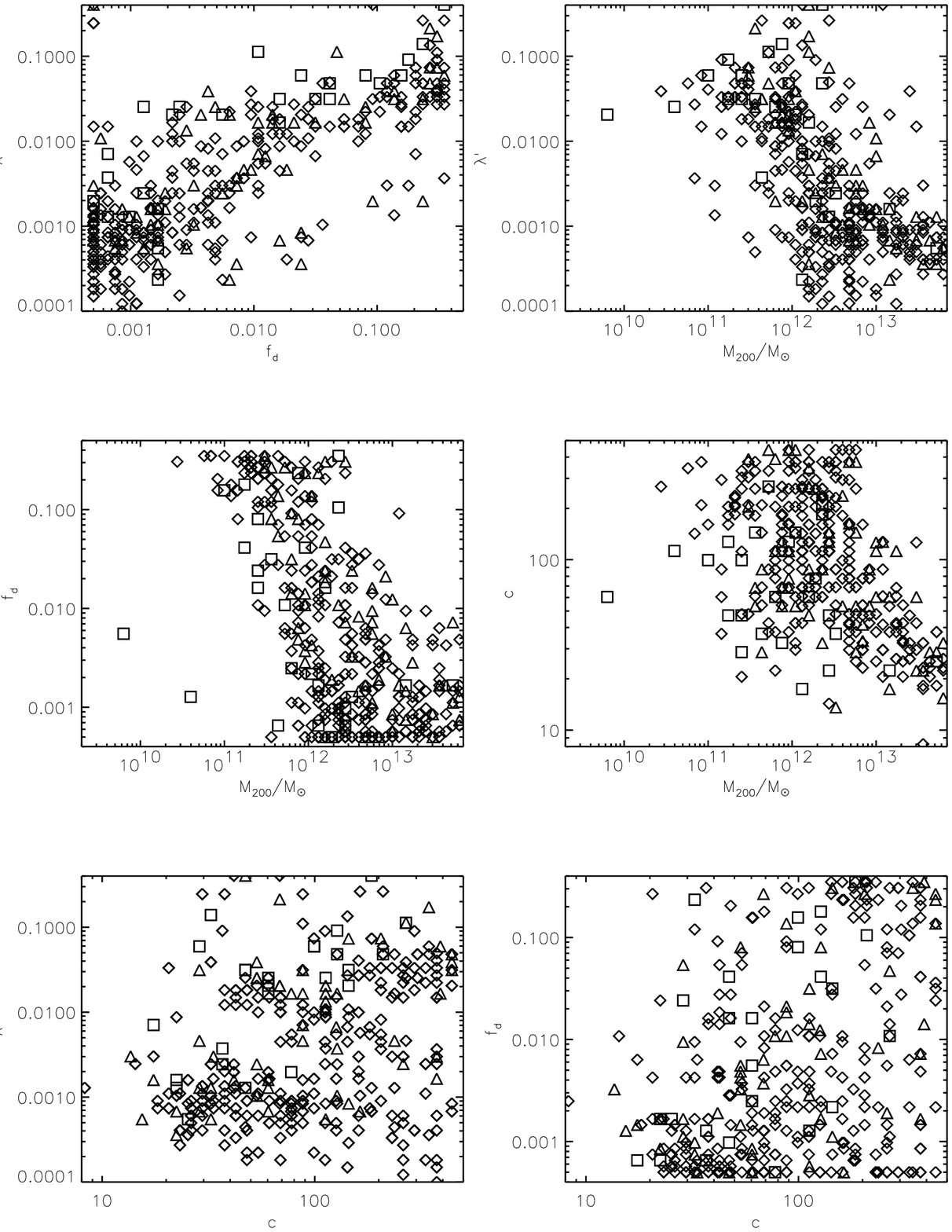}
\caption{Same as fig.~\ref{fig:hist} but for the model with disks inside
  pseudo-isothermal dark halos.  Note that the correlations between $\lambda$
  and $f_d$ and between $M_{200}$ and $f_d$ are very similar to those found
  for the NFW model and to those found for the three parameters fit. Note in
  the $\lambda'-f_d$ panel that LSBs have higher values of $\lambda'$ than
  HSBs for a given value of $f_d$. Also, in the $c-M_{200}$ panel, LSBs have
  lower values of $c$ than HSBs for a given mass.}
\label{fig:histpseu}
\end{figure*}

\section*{Appendix II: Fit to 4 parameter models}

In this Appendix we explore how a model with four free parameters ($M_{200}$,
$f_d$, $\lambda$ and $c$), i.e. without fixing $R_d$ from observations, fares
against the three free parameters model used in the paper. This is interesting
for two reasons: it allows us to estimate how much worse the parameter
estimation becomes when adding an extra free parameter and allows us to assess
the robustness of the zones of avoidance. As in the paper we study both a NFW
and a pseudo-isothermal dark matter halo. We also include the LSB sample and
show that the trends for parameter correlations for LSBs are the same as for
the Courteau and PW samples (HSBs).

In this case, for each galaxy rotation curve we explore the entire likelihood
surface within the same limits as the three free parameter model and $0.001 <
\lambda' < 0.5$.

In fig.~\ref{fig:hist} and \ref{fig:histpseu} we show the position of each
galaxy in all six projections of the four dimensional space
$(f_{d},M_{200},\lambda^{\prime},c)$, for the NFW and pseudo-isothermal models
respectively. Diamonds correspond to galaxies from the \citet{Courteau97}
sample, triangles to galaxies from \citet{PW00} and squares to the LSB
galaxies from \citet*{MRB01}.

First we discuss the case of the NFW profile (fig.~\ref{fig:hist}). The same
correlations and zone of avoidance are present as in the three free parameter
model; for 144 galaxies out of the 400 no fit was found within the boundaries.
The upper-left panel of fig.~\ref{fig:hist} shows the same zone of avoidance
(low values of $f_d$ and high values of $\lambda^{\prime}$ are avoided) as
before and also a similar correlation between $f_d$ and $\lambda^{\prime}$
although not as clear as the one in fig.~\ref{fig:nfwall}.

The change in the resulting distribution of recovered parameters can be seen
clearly in the plane $c-M_{200}$. Here a large group of galaxies have
recovered values of $c$ and $M_{200}$ in excellent agreement with those found
in fig.~\ref{fig:nfwall}. On the other hand, a clump with unphysical values of
$c < 1$ is clearly visible. They correspond to galaxies with unphysically high
values of $f_{d}$ (see the corresponding clump in the $M_{200}-f_{d}$ plane).
About 25\% of the rotation curves favours this different solution. These are
low spatial resolution rotation curves: their quality is comparable to that of
f583-4 shown in fig.~\ref{fig:plot3rc}.

Note that in all 3 cases, the degeneracies between parameters run in almost
exactly the same direction as the correlations recovered from the data. These
correlations may therefore be the result of parameter degeneracies if errors
in the rotation curve have been underestimated.

For the four free parameter case considered here the recovered $M/L$ are
slightly broader than for the three free parameter case.  This indicates that
for this case we are still recovering correctly the mass of the disk.  We have
computed the $M/L$ for the eight systems in the LSB sample with reliable
photometry. For these systems, $M/L=0.1 - 2.5$ in the $B$ band for both the
NFW and pseudo-isothermal profile, in fair agreement with the predictions from
synthetic stellar population models.

The recovered parameters for the pseudo-isothermal dark matter halo are shown
in fig.~\ref{fig:histpseu}. Also in this case the zones of avoidance are
preserved and the correlations among the different parameters are remarkably
similar to those obtained from the three free parameter model. In this case
for 141 galaxies out of 400, no fit was found within the boundaries for the
parameters.

From fig.~\ref{fig:histpseu} it can be seen that LSB galaxies tend to have
higher values of $\lambda'$ for a given $f_d$ and lower values of $c$ for a
given mass than HSB galaxies. A similar trend occurs for the NFW profile.

\begin{figure}
\includegraphics[width=8.5cm,height=12cm]{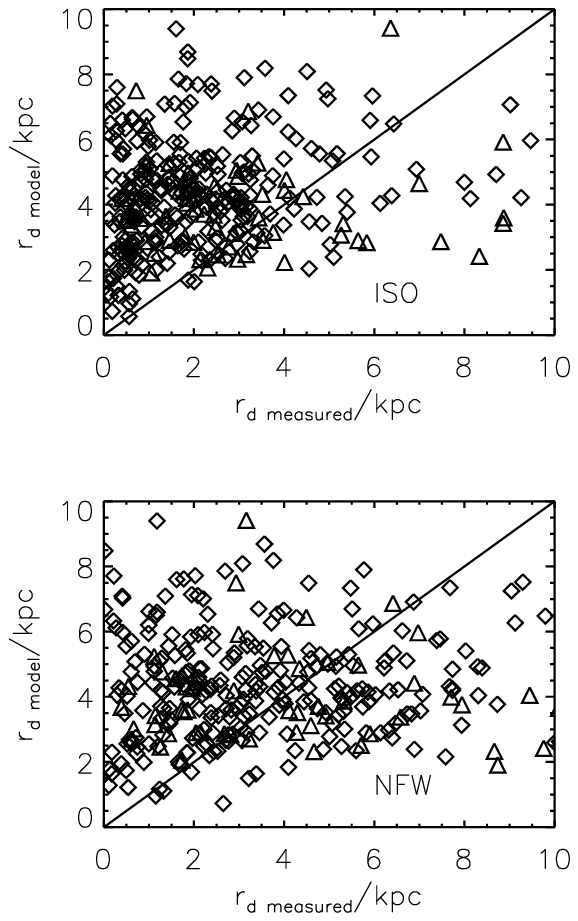}
\caption{Comparison of the recovered scale-length of the disk, when using 
  a four free parameter model, for the pseudo-isothermal model (top panel) and
  the NFW model (bottom panel) with the observed stellar scale-lengths. The
  disk scale-lengths are very poorly constrained from the rotation curve
  alone.}
\label{fig:rdvsrd}
\end{figure}

Figure~\ref{fig:rdvsrd} shows the recovered values of $R_d$ in this case
plotted against the measured values from the photometry for the two profiles.
We see that the disk scale length is very poorly constrained from the rotation
curves alone. In particular, for the pseudo-isothermal case the estimated disk
scale-length is systematically several times larger than the true disk scale
length. This corresponds to inserting a constant surface density sheet over
the extent of the rotation curve.

\section*{Appendix III: Degeneracies in the model}

\label{section:error_analysis}
Similar modelling to that performed here, but with HI rotation curves, has
shown that there are serious model degeneracies (e.g., \citet*{vanBBS01}). For
example, it is common to expect a degeneracy between $M_{200}$ and $c$ (see
e.g., \citet*{BMR01}). In this section we argue that the main reason for this
is the poor spatial resolution of the HI curves and not the model itself. To
demonstrate this we will explore the likelihood surfaces of the 400 NFW fits
(same conclusions derived in this section apply to the pseudoisothermal fits).
We choose to do this using fits to 4 parameters to demonstrate that even with
such a large number of free parameters degeneracies are removed for
high-resolution rotation curves. Obviously, for the case of only 3 free
parameters, possible degeneracies will be even more restricted.

\begin{figure}
\includegraphics[width=8.5cm]{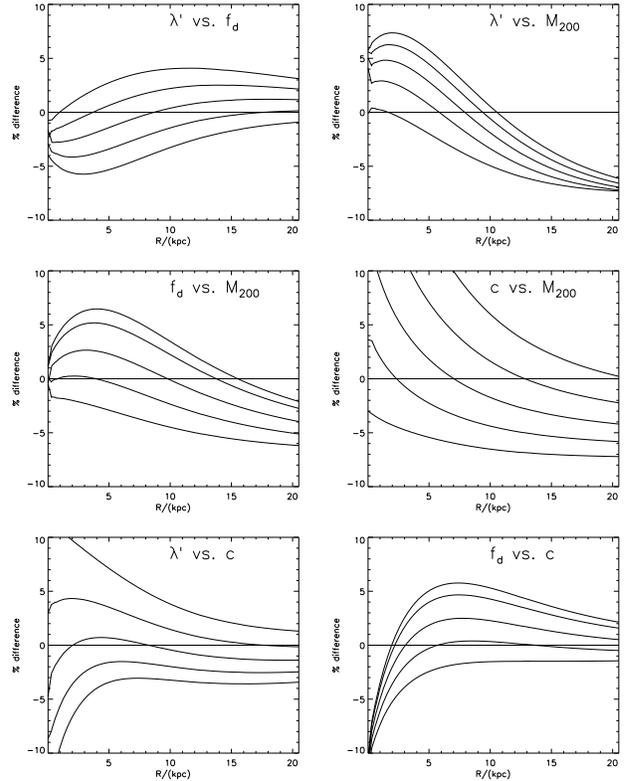}
\caption{The percentage difference between two rotation curves for which two
  parameters are changed at the time while the other two are kept constant in
  a fiducial model with ($M_{200}$, $\lambda'$, $f_d$, $c$)=($1 \times
  10^{11}$, $0.05$, $0.01$, $10$). The different curves correspond to
  variations of 5, 10, 15, 20 and 25\% in the model parameters. The two
  parameters that are changed are shown on the top of each panel. The fiducial
  model is shown as a solid horizontal line. If the rotation curve is known
  with less accuracy than the typical variation among parameters, these will
  be degenerated.}
\label{fig:degenerot}
\end{figure}

First, it is useful to look at how the model behaves under changes of
parameter values. To illustrate this we choose a fiducial model with
($M_{200}, \lambda^{\prime}, f_d, c$)=($1 \times 10^{11}$ M$_{\odot}$, 0.05,
0.01, 10) and explore some of the possible degeneracies present in the
rotation curve.  To do this we keep two parameters fixed and allow the other
two to change from their fiducial value.  This results in six possible
combinations. There will be degeneracy if in any of these combinations the two
rotation curves are indistinguishable. Fig.~\ref{fig:degenerot} shows the
percentage difference of the rotation curve so obtained with the fiducial
model. Note that if the rotation curve is not known to better than 5\%, then
there are clear degeneracies in the $\lambda^{\prime}-f_d$,
$\lambda^{\prime}-M_{200}$ and $f_d-M_{200}$ planes. On the other hand, in the
other planes there are no degeneracies if the rotation curve is
well-constrained below 2 disk scale-lengths. Therefore, if the rotation curve
is known to better than 5\%, with significant spatial coverage, degeneracies
among the parameters should not arise.  Of the various parameter combinations,
the most degenerate pair is $\lambda^{\prime}-f_d$.

\begin{figure}
\includegraphics[width=8.5cm]{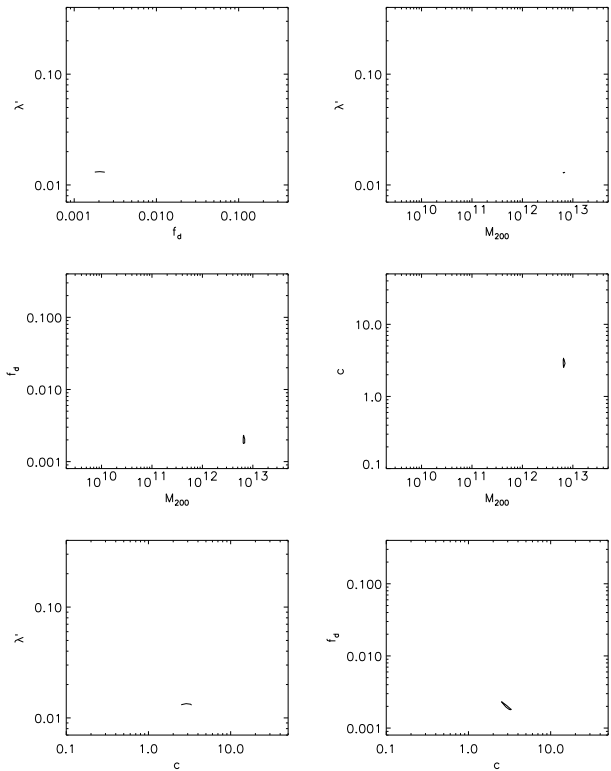}
\caption{The 95.4\% confidence contours for galaxy 11846 in the Courteau
  sample for a pair of parameters marginalizing over the other two. Note the
  small size of the confidence regions due to the good spatial sampling of the
  rotation curve.}
\label{fig:11846}
\end{figure}

We now turn our attention to the actual data. We first choose a good example
to illustrate how well degeneracies can be broken. We choose galaxy 11846 from
the \citet{Courteau97} sample. Fig.~\ref{fig:11846} shows the 95.4\%
confidence contour for two parameters having been marginalized, i.e.
integrated the likelihood (${\cal L}=\exp[-\chi^2/2]$), on the remaining
parameters (note that these quoted confidence intervals should not be
over-interpreted: the true (larger) uncertainty is likely to be dominated by
systematic errors and inadequacies in the model).  The confidence contours are
indeed rather small, implying errors in the recovered parameters of only about
20\%. From a Fisher matrix analysis (see section \ref{sec:fisher}) we find
that about 20\% of the rotation curves in our sample have the above quality,
60\% have $\sim 50\%$ errors in the recovered parameters. Only 20\% have much
larger error bars, with much poorer constraints on the recovered parameters;
one of these galaxies is f583-4.

\begin{figure}
\includegraphics[width=8.5cm]{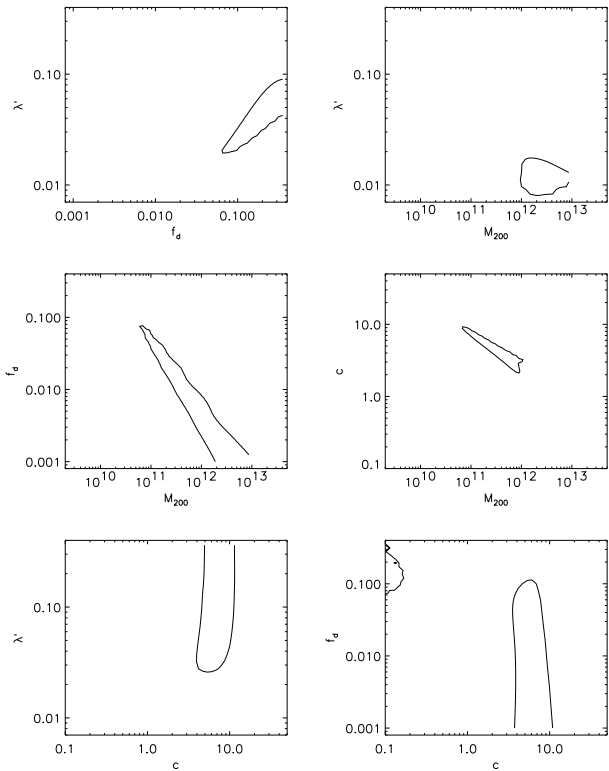}
\caption{Same as fig.~\ref{fig:11846} but the galaxy f583-4 from the de Block
  et al. (2001) sample which has a poor spatially resolved rotation curve. In
  this case we show the 68.5\% confidence contours. Due to the poor spatial
  resolution the confidence contours are much larger in this case. It is also
  useful to see that the regions of degeneracy are not inside the avoidance
  regions.}
\label{fig:f583-4}
\end{figure}

The case of f583-4 is extremely useful to illustrate the large degeneracies
that arise due to the lack of spatial resolution in the rotation curve. This
has only 9 points with large error bars. Fig.~\ref{fig:f583-4} shows the
68.5\% confidence contours. These are large and broad, and the parameters are
very poorly constrained. This should not come as a surprise since one can
hardly expect to constrain well 4 parameters with only 9 points in view of the
results from fig.~\ref{fig:degenerot}.  We stress here that 80\% of the
galaxies in the sample have much better resolution rotation curves thus the
recovered parameters are much better constrained.

We explore the dependence of parameters on the spatial resolution of
the rotation curve by investigating in further detail e376g2 from the
\citet{PW00} sample. We perform the following manipulations to
e376g2: first we have removed all points beyond 10 kpc (no-flat), 
to produce a rotation curve without the flat part. Next, to the original
rotation curve we remove the inner points below a radius of 3
kpc, so there is no rising part of the rotation curve (no-r). 

We also select a random subsample, only $1/3$ of the original points, from the
rotation curve, thus degrading spatial resolution (low-res). Finally, we fit
independently the approaching and receding parts of the rotation curve. This
gives us a handle on the effect of asymmetries in the rotation curve. The
values for the parameters are shown in table~\ref{table:table}. Also shown are
the values for the full fit (all) with errors obtained from marginalizing over
the other three parameters. The variation in the recovered values of
$\lambda^{\prime}$ is very small, while variations of up to 100\% occur for
the other parameters.  The larger error in the recovered values happens for
the low-res case, i.e.  when the spatial resolution of the rotation curve is
decreased. It is interesting to see how the different parameters are affected.
The largest error for $M_{200}$ is for the low-res case, while the largest
error for $c$ occurs when only one part of the rotation curve is used.  For
$f_d$ the largest deviation occurs in the low-res case.

\begin{figure*}
\includegraphics[width=16cm,height=5cm]{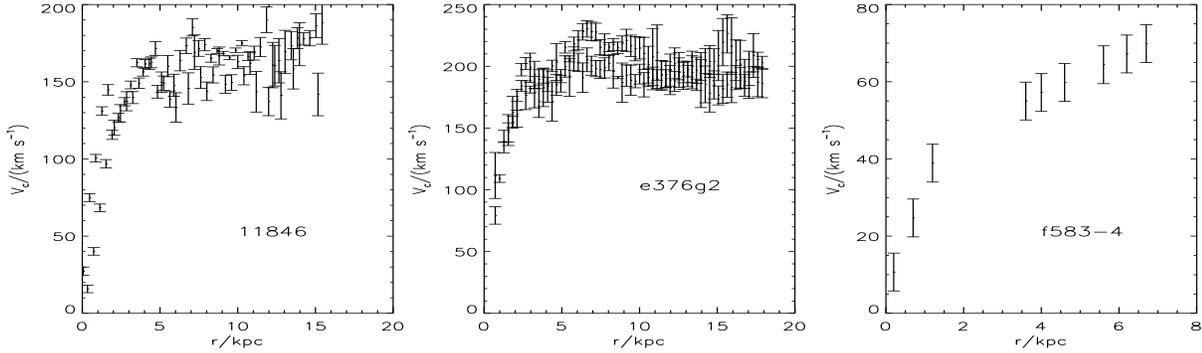}
\caption{Three examples of rotation curves. The one on the left panel is from
  the Courteau sample, the one in the middle is from the PW sample and the one
  on the right panel from the LSB sample.}
\label{fig:plot3rc}
\end{figure*}

The three rotation curves used in the above example are shown in
fig.~\ref{fig:plot3rc}.

\begin{table}
\begin{tabular}{ccccccc}
e376g2 & all & no-flat & no-r & low-res & left & right \\
\hline
$M_{200}$ & $6 \pm 3$ & 5 & 6 & 4 & 10 & 7 \\ 
$c$ & 4 $\pm 1$ & 3 & 4 & 4 & 4 & 2 \\
$f_d \times 10^{-3} $ & 4 $\pm 1$ & 6 & 4 & 8 & 2 & 8 \\
$\lambda^{\prime} \times 10^{-2}$ & 13 $\pm 5$ & 13 & 13 & 15 & 12 & 13\\
\hline
\end{tabular}
\caption{Values of recovered parameters from the rotation curve of e376g2
  after having performed several manipulations to its rotation curve (see
  text). The mass in units of $10^{12}$ M$_{\odot}$.}
\label{table:table}
\end{table}

\subsection{Fisher matrix analysis}
\label{sec:fisher}

When the galaxy sample is large, it might be inefficient to explore the whole
likelihood surface as we have done in this case and maybe more important, it
might not be possible in the case of very large datasets. Therefore it is
useful to use a fast algorithm to estimate the error and, especially, the
possible degeneracies. As an alternative to the exact study presented above,
and as an illustration for a method that can be suitable for very large data
sets, we apply the Fisher matrix analysis to our data set.

We performed a Fisher matrix analysis of the likelihood function
${\cal L}$, where $\chi^2=-2 \ln({\cal L})$. Recall that for $M$
parameters, the marginal error on each of the parameters can be
obtained, under the assumption of Gaussian likelihood, from the Fisher
information matrix
\begin{equation}
F_{ij} \equiv -\left\langle {\partial^2\ln{\cal L}\over \partial
\theta_i \partial\theta_j}\right\rangle
\end{equation}
where $i$ and $j$ run from 1 to $M$ (in our case $M=4$) and  the angle brackets
denotes the averaging over many realizations.
The marginal error is then
\begin{equation}
\sigma_i = \sqrt{(F^{-1})_{ii}}.
\end{equation}
Since our best fit parameters are those that maximize the likelihood, a Taylor
expansion of $\ln {\cal L}$ around $\bbtheta_0$ gives
\begin{equation}
\ln {\cal L}(\bbtheta_0+\Delta\bbtheta) \simeq \ln {\cal L}_0 + {1\over
2}{\partial^2\ln{\cal L}\over \partial \theta_i
\theta_j}\Delta\theta_i \Delta\theta_j.
\end{equation}
The  elements of the Fisher matrix can be estimated as:
\begin{eqnarray}
F_{ii}\!\!& & \!\!\!\simeq \\\nonumber
&-\!\!&\!\!\!{1\over \Delta\theta_i^2} \left[\ln {\cal L}(\bbtheta_0 +
\Delta\theta_i {\bf e}_i) +  \ln {\cal L}(\bbtheta_0 -
\Delta\theta_i {\bf e}_i) - 2 \ln {\cal L}_0\right].
\label{Fii}
\end{eqnarray}
and
\begin{eqnarray}
F_{ij} & \simeq & {-1\over 2 \Delta\theta_i \Delta\theta_j} \times \\\nonumber
& &\left[\right.lnL(\bbtheta_0 + \Delta\theta_i {\bf e}_i + \Delta\theta_j {\bf e}_j)+
\\\nonumber
& &\lnL(\bbtheta_0 - \Delta\theta_i {\bf e}_i -
\Delta\theta_j {\bf e}_j)- \\\nonumber
& &\lnL(\bbtheta_0 - \Delta\theta_i {\bf e}_i +
\Delta\theta_j {\bf e}_j)- \\\nonumber
& &\lnL(\bbtheta_0 + \Delta\theta_i {\bf e}_i -
\Delta\theta_j {\bf e}_j)
\left .\right].
\label{Fij}
\end{eqnarray}

Furthermore, since the Fisher matrix is symmetric it can be
diagonalized ($F= U \Lambda U^{T}$, where $\Lambda$ is a diagonal
matrix with elements the eigenvectors $\lambda_1,...,\lambda_M$) and a
new orthogonal set of variables built such that $N=U^{T} O$, where $O$
is the original set of variables ($\lambda$, $c$, $f_d$ and $M_{200}$
in our case). Thus each row of $U^{T}$ gives the amount by which the
original variables are mixed and therefore how degeneracies arise
among them.

We have performed the above Fisher analysis on the 400 rotation curves, both
for the NFW and the pseudo-isothermal profile. Our conclusions regarding
degeneracies are in agreement with the findings of the full analysis performed
on galaxies 11846 and f583-4 and illustrated in figures \ref{fig:11846} and
\ref{fig:f583-4}.

\end{document}